\documentclass[11pt]{article}

\usepackage{mathptmx}      
\usepackage{graphicx}
\usepackage{eucal}
\usepackage{mathrsfs}
\usepackage{amsfonts} 
\usepackage{amssymb}
\usepackage{amsthm}
\usepackage{amsmath} 
\usepackage{tabularx}
\usepackage{url}
\usepackage{enumerate}
\usepackage{multirow}
\usepackage{bigstrut}

\usepackage{pseudocode}
\usepackage{float}
\newfloat{algo}{h}{alg}[section]
\floatname{algo}{Algorithm}
\def\algocapt#1{\addcontentsline{alg}{subsection}%
		{\protect\numberline{\csname thepseudocode\endcsname}%
		{\ignorespaces #1}}}

\newcommand{\R}{{\mathbb R}}
\newcommand{\N}{{\mathbb N}}
\newcommand{\argmin}{{\operatorname{argmin}\,}}
\newcommand{\abs}[1]{\left\vert#1\right\vert}

\newtheorem{theorem}{Theorem}[section]
 
\newtheorem{lemma}[theorem]{Lemma}

\theoremstyle{definition}

 \theoremstyle{remark}

\newtheorem{example}[theorem]{Example}

\begin{document}
\title{Indexing Schemes for Similarity Search In Datasets of Short Protein Fragments
}

\author{Aleksandar Stojmirovi\'{c}\footnote{Present address: National Center for Biotechnology Information, National Library of Medicine, National Institutes of Health, Bethesda, MD 20894, United States of America.} \ and \ Vladimir Pestov 
\\[.5cm]
 Department of Mathematics and Statistics, University of Ottawa \\
 585 King Edward Avenue \\
 Ottawa, Ontario, Canada. K1N 6N5 \\
 Tel.: +1-613-562 5800 ext. 3523 \\
 Fax: +1-613-562 5776\\[3mm]
 {\tt stojmira@ncbi.nlm.nih.gov, vpest283@uottawa.ca}
}

\maketitle
\begin{abstract}
We propose a family of very efficient hierarchical indexing schemes for ungapped, score matrix-based similarity search in large datasets of
short (4-12 amino acid) protein fragments. This type of similarity search has importance in both providing a building block to more complex algorithms and for possible use in direct biological investigations where datasets are of the order of 60 million objects. 
Our scheme is based on the internal geometry of the 
amino acid alphabet and performs exceptionally well, for example outputting 100 nearest neighbours to any possible fragment of length 10 after scanning on average less than one per cent of the entire dataset.
\\[2mm]
\end{abstract}


\section{Introduction}

\begin{table}[t!]
\begin{center}
\begin{tabular}{|l|l|}
\hline
 $\Sigma$ & Amino acid alphabet.\\ 
 $m$ &  Fragment length.\\
 $X$ &  Fragment dataset. \\
 $n$ &  Size of $X$; usually not known exactly beforehand.\\
 $f$ & Query function on $\Sigma^m$: $f(x)=\sum_{i=0}^{m-1}f_i(x_i)$.\\
 $f_i$ & $i$-th component of the query function.\\  
 $Q^\text{rng}_f(\varepsilon)$ & Range query with respect to $f$ with threshold $\varepsilon$: all $y\in X$ such that $f(y)\leq\varepsilon$.\\
 $Q^\text{$k$NN}_f(k)$ & $k$-nearest neighbours query with respect to $f$: $k$ points from  $X$ with the smallest values of $f$.\\
 $s$ & Similarity score on $\Sigma^m$: $s(x,y)=\sum_{i=0}^{m-1}S(x_i,y_i)$\\
 $S$ & Similarity score matrix over $\Sigma$: a map $S:\Sigma\times\Sigma\to\R$.\\
 $d$ & A quasi-metric distance (over $\Sigma^m$ but also used in general context).\\
 $D$ & Distance matrix over $\Sigma$: a map $D:\Sigma\times\Sigma\to\R_+$.\\
 $\rho$ & A metric symmetrization of the quasi-metric $d$.\\
$w$ & Weight function for a co-weightable quasi-metric space.\\
 $\Omega(z)$ & A fibre: subset of $\Omega$ with constant weight. \\
  $\Gamma_i$ & Reduced alphabet at the $i$-th position.\\ 
 $\pi_i$ & Projection $\Sigma\to\Gamma_j$ at the $i$-th position.\\
 $\pi$ & Projection function: maps each fragment into its bin.\\ 
 $\gamma$ & A letter in a reduced alphabet.\\
 $\mathscr{B}$ & Set of all bins.\\ 
 $N$ &  Total number of bins: $N=\prod_{i=0}^{m-1}\abs{\Gamma_i}$.\\ 
 $\xi_i$ & Integer value of a letter of the reduced alphabet at the $j$-th position.\\
 $r$ &  Bin ranking function: index into the $bin$ array.\\
 $u$ &  Index of a bin: $u=r(B)$ where $B$ is a bin.\\
 $F_i$ & Lower bound to $f_i$ over the reduced alphabet at position $i$: $F_i(\gamma)=\min\{f_i(a)\ |\ a\in\gamma\}$.\\
 $F$ & Lower bound to the query function over bins: $F(B)=\sum_{i=0}^{m-1}F_i(B_i)$\\
\hline
\end{tabular}
\end{center}
\caption{Summary of symbols.}\label{tbl:FSvars}
\end{table}

Indexing of biological sequences for fast similarity search has received a lot of attention in recent years \cite{HuAtIr01,Hu04,Buhler01,GiWaWaVo00,MXSM03,Kent:2002,TCOT03}. While many schemes were proposed for indexing DNA sequences, much less effort has been spent on protein sequences. This is understandable because the DNA datasets are many times larger and offer potentially more information.  However, similarity search in protein datasets differs in a few important aspects from the corresponding DNA based search: the protein alphabet is larger and the similarities between any two individual amino acids are not all identical but are given instead by a score matrix such as a member of the PAM \cite{Dayhoff:1978} or BLOSUM \cite{Henikoff:1992} family.

In this paper we focus not on the whole protein sequences but on their fragments. We take as our datapoints the set of all peptide fragments of fixed length (also known as \emph{q-grams} in the more general context of string matching \cite{Ukkonen92a}) taken from a protein sequence database. Our similarity measure is global: the similarity score is given by the sum of the scores of amino acids at the same position, that is, by extension of a similarity measure on the amino acid alphabet. As the fragments we investigate are quite short (length 4-12), we do not allow gaps.

Our motivation is twofold. Firstly, there are reasons to believe that similarity search in peptide fragments can become an important tool for function discovery. Short peptide motifs can form parts of larger functional domains and often have important physiological function on their own \cite{martinserano01,kitts03}. We are interested in large-scale searches for biologically `interesting' short peptide motifs. In that case, a search algorithm becomes an important subroutine, to be followed by filtering of search hits using additional criteria, since raw similarity score usually does not provide sufficient grounds for assumption of sequence homology for short protein fragments.

Secondly, sets of short peptide fragments provide a very convenient object of study in the context of designing general indexing schemes for similarity search. Those datasets are quite sizable but not excessively large and are equipped with a simple and computationally cheap similarity measure. We have already used an early implementation of one of the indexing schemes in a
family described here (FSIndex) as an illustration for general indexing concepts \cite{PeSt06}. 

The family of indexing schemes we present here, called FSIndex ('Functional Sequence Index'), is based on the idea that amino acids can be grouped according to their chemical properties and that these groupings are, albeit imperfectly, reflected in almost all similarity measures used in practice. Precomputing distances (or more generally, query functions) from amino acids to clusters allows efficient computation of the lower bounds for the distance between query and database fragments and pruning of a large proportion of the dataset that can be certified not to belong to a similarity-based query. Unlike with the most other proposed indexing schemes, the hierarchical traversal of the search space is performed using an implicit, combinatorially generated tree, so that only the data points and the clustering information need to be stored.

\section{Background}

\subsection{Sets of protein fragments}

We model protein fragments of fixed length $m$ as strings of length $m$ generated by the standard 20 amino acid alphabet $\Sigma$ and denote the set of all such strings by $\Sigma^m$. We write $x=x_0x_1\ldots x_{m-1}$ for each $x\in\Sigma^m$. Obviously, the same approach can be used for any sets of strings of fixed length generated by a finite alphabet. A protein fragment dataset or instance (which we will usually denote by $X$) is therefore a subset of $\Sigma^m$.

To describe similarity-based queries, we define a similarity measure on $\Sigma^m$ as a function $f:\Sigma^m\to\R$ such that $f(x)=\sum_{i=0}^m f_i(x_i)$, where $f_i$ are functions $\Sigma\to\R$ for all $i\in I_m$ ($I_m=\{0,1,\ldots,m-1\}$). A range query based on $f$ with range $\varepsilon$, denoted $Q^\text{rng}_f(\varepsilon)$, retrieves all points $x\in X$ such that $f(x)\leq\varepsilon$. A $k$NN ($k$ nearest neighbours) query, denoted $Q^\text{$k$NN}_f(k)$, retrieves $k$ points from $X$ that have the smallest value of $f$. We call the queries based on $f$ the half-plane or valuation queries.

The above setting generalises the similarity queries based on a dissimilarity measure, that is, a distance function. Indeed, if we have a distance $d:\Sigma^m\times\Sigma^m\to\R_+$ and a point $\omega\in\Sigma^m$, retrieving all points $x\in X$ such that $d(\omega,x)\leq \varepsilon$ is equivalent to setting up a range query $Q^\text{rng}_f(\varepsilon)$ where $f(x)=d(\omega,x)$. While in the context of this paper we will mostly use distances to generate queries, there are practically relevant cases where the more general formulation is desirable.

In the context of biological sequences, similarity scores that are large and positive for similar sequences and small and negative for dissimilar ones, are most commonly used. For sequences of arbitrary lengths, Needleman-Wunsch \cite{NW70} and Smith-Waterman \cite{SW81} algorithms are used to compute global and local similarity scores respectively, depending on the score matrix $S$ over the alphabet and on gap penalties. Both algorithms take the time at least quadratic in the length of sequence and can also be used to compare the sequences of fixed length. 

Our choice however, is an ungapped similarity measure, depending solely on the score matrix, because gaps rarely appear in alignments of fragments of short length (the penalty for opening gaps is usually set to be large compared to penalties for substitutions of letters). For two fragments $x,y\in\Sigma^m$, we have $s(x,y)=\sum_{i=0}^{m-1} S(x_i,y_i)$ and a range query centred at $\omega\in\Sigma^m$ with cutoff threshold $t$ retrieves all points $x\in X$ such that $s(\omega,x)\geq t$. The fragment similarity measure described above has a significant advantage in that it can be computed in time linear in $m$, the length of fragments.

Position Specific Score Matrices (PSSMs) \cite{Gribskov:1987} provide a generalisation of score matrix based similarity scores that produce more accurate search results. Instead of using the same score matrix $S$ at each position $i\in I_m$, a different score function is used. PSSMs are constructed from sets of aligned sequences classified according to some property and are used primarily to detect similarities that cannot be found using score matrices. An iterative procedure is very commonly used \cite{altschul97gapped}: a `seed' query sequence is provided and the first search is performed using a score matrix. The results of that search are then used to construct a PSSM and another search is performed based on it. The search can be further refined as necessary by using the results of one search to construct the PSSM used in the subsequent query.

The importance of PSSMs in biological applications motivates our choice of the similarity measure on $\Sigma^m$ as a function of the form $f(x)=\sum_{i=0}^m f_i(x_i)$ since a PSSM cannot be expressed as a standard score matrix. Here the functions $f_i$ are given by the columns of the PSSM.

\subsection{Quasi-metrics}

Let $\Omega$ be a set. A map of two arguments, $d: \Omega\times \Omega\rightarrow \R_+$, is called a quasi-metric if it satisfies the properties: (i) for all $x,y\in \Omega$, $d(x,y)=d(y,x)=0 \Leftrightarrow x=y$, and (ii) for all $ x,y,z\in \Omega$, \ $d(x,z) \leq d(x,y)+ d(y,z)$ (triangle inequality). It is called a metric if in addition it satisfies $d(x,y)=d(y,x)$ for all $x,y\in \Omega$ (symmetry). Quasi-metrics form an area of active research in general topology \cite{Ku01} but the knowledge of the geometry associated with them is still relatively scarce. 

\begin{figure*}[t]
\begin{center}
\begin{tabular}[t]{lcr}
\scalebox{0.8}{\includegraphics{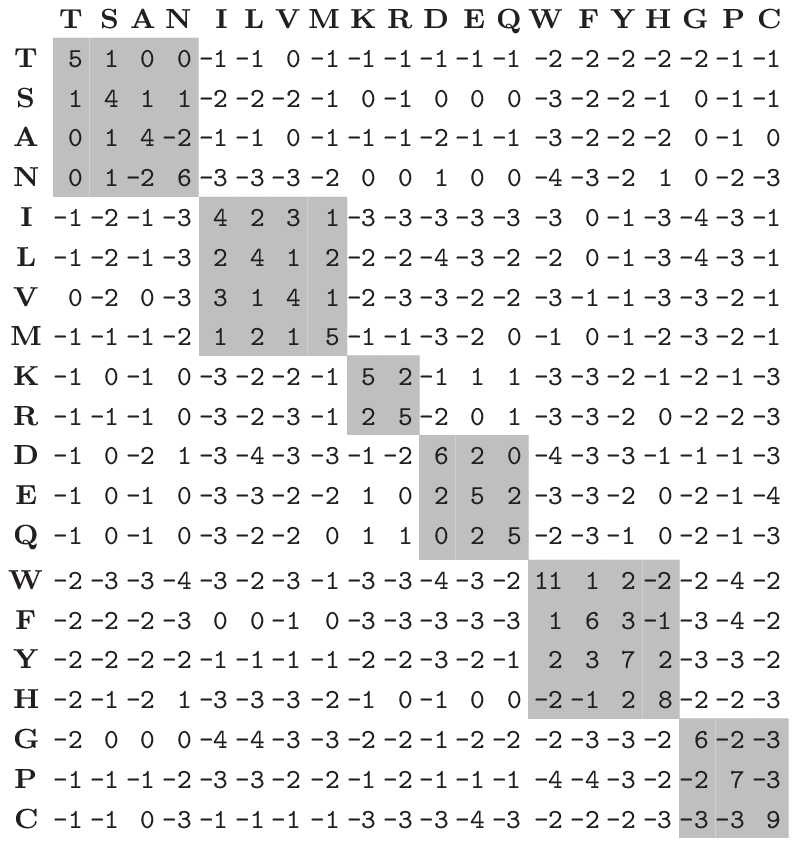}} &
\scalebox{0.8}{\includegraphics{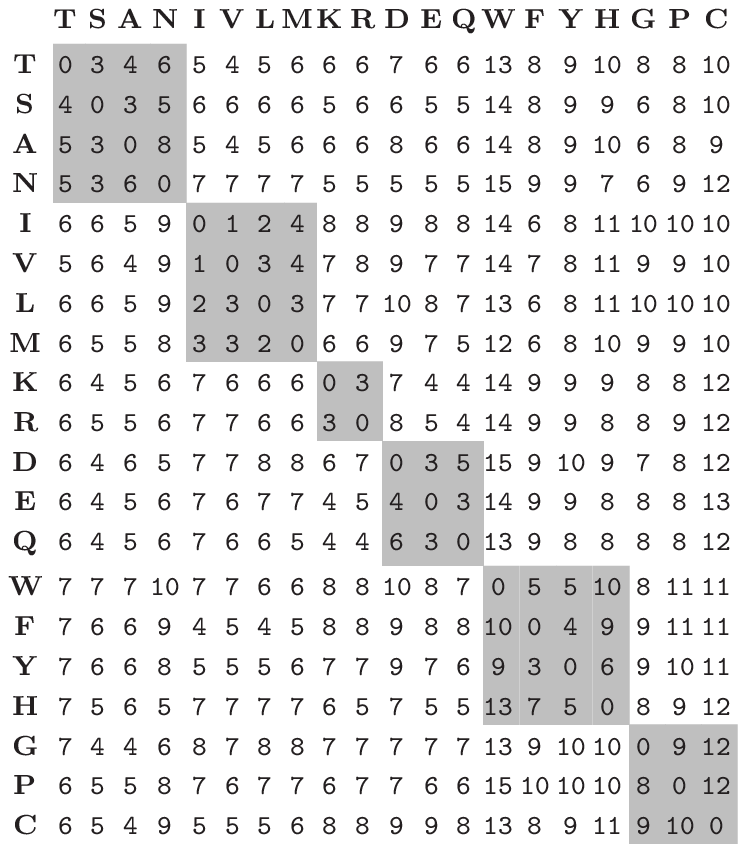}}\\ &
\end{tabular}
\caption{BLOSUM62 similarities (left) and quasi-metric distances (right). Distances within members of an alphabet partition used for constructing an index for fragments of length 9 used in experiments are grayed.}\label{fig:blosum62qd}
\end{center}
\end{figure*}

Quasi-metrics are relevant to our investigation because of their relation to similarity scores. It can be verified \cite{AS2004}
that most of the matrices from the BLOSUM \cite{Henikoff:1992} family, the most widely used family of score matrices over the amino acids, restricted to the standard amino acid alphabet, produce a quasi-metric under the transformation $D(x,y) = S(x,x) - S(x,y)$. In particular, the matrices BLOSUM 45, 50 60, 62 (Figure \ref{fig:blosum62qd}), 65, 80, 90 and 100 do so while BLOSUM 30, 35, 40, 55, 70 and 75 do not. As the triangle inequality is violated only once for BLOSUM 55, 70 and 75, we suspect that its violation may be due to round-off error. No member of the also widely used PAM \cite{Dayhoff:1978} family of score matrices can be transformed into quasi-metrics in this way but the number of failures of the triangle inequality is small for at least some of them, indicating that they can be closely approximated by quasi-metrics. 

The above transformation canonically extends to fragments: the distance function given by $d(x,y) = s(x,x) - s(x,y)$ is then also a quasi-metric. Its main significance is that every range query centred at $\omega$ with threshold $t$ based on similarity score $s$ coincides with the range query centred at $\omega$ with respect to the quasi-metric $d$ of radius $\varepsilon=s(\omega,\omega)-t$. 

\begin{example}\label{ex:1}
Consider a space of fragments of length 3 from the alphabet $\Sigma=\{a,b,c,d\}$ where the similarity score matrix $S$ is given in Figure \ref{fig:examplemat} (a). Using the transformation $D(\sigma,\tau) = S(\sigma,\sigma) - S(\sigma,\tau)$ we obtain the distance matrix $D$ (Figure \ref{fig:examplemat} (b)).

\begin{figure}[!h] 
\begin{center}
\begin{tabular}[t]{lcr}
\multicolumn{1}{l}{\mbox{\bf (a)}} & \rule{2cm}{0pt}&
               \multicolumn{1}{l}{\mbox{\bf (b)}}  \\ 
\begin{tabular}{|l|rrrr|}
\hline
$S$ &  a &  b &  c &  d\\ \hline
a&  5 & -3 &  2 & -2\\
b& -3 &  5 & -4 &  3\\
c&  2 & -4 &  6 & -4\\
d& -2 &  3 & -4 &  6\\
\hline
\end{tabular} & &
\begin{tabular}{|l|rrrr|}
\hline
$D$ &  a &  b &  c &  d\\ \hline
a&  0 & 8 &  3 & 7\\
b& 8 &  0 & 9 &  2\\
c&  4 & 10 &  0 & 10\\
d& 8 &  3 & 10 &  0\\
\hline
\end{tabular}\\
\end{tabular}
\caption{An example of a similarity score \textbf{(a)} and the corresponding distance matrix \textbf{(b)}.}\label{fig:examplemat}
\end{center}
\end{figure}

It can be verified that $D$ is indeed a quasi-metric on $\Sigma$. Consider the fragments $x=abd$, $y=cad$ and $z=cbb$. Using similarity scores, we have $s(x,x)=S(a,a)+S(b,b)+S(d,d)=16$, $s(x,y)=S(a,c)+S(b,a)+S(d,d)=5$, $s(x,z)=S(a,c)+S(b,b)+S(d,b)=10$, while using distances, $d(x,y)=D(a,c)+D(b,a)+D(d,d)=11$ and $d(x,z)=D(a,c)+D(b,b)+D(d,b)=6$. We can verify that $d(x,y)=s(x,x)-s(x,y)$ and $d(x,z)=s(x,x)-s(x,z)$.

Now consider a range similarity query about $x$ with the threshold $9$, that is, the query retrieving all points $u$ such that $s(x,u)\geq 9$. The point $z$ above belongs to this query while $y$ does not. Since $s(x,x)=16$, we can transform this query to a query with respect to $d$ of radius $16-9=7$. Using our general notation we represent this query by $Q^\text{rng}_f(7)$ where $f(u)=d(x,u)$.
\end{example}

If a similarity score matrix is symmetric (as all of BLOSUM and PAM matrices are), the corresponding quasi-metric is co-weightable \cite{KuVa94}, that is, it satisfies $d(x,y) + w(y) = d(y,x) + w(x)$ for all $x,y\in\Sigma^m$, where $w(x)=s(x,x)$ is called the weight function. Therefore, we can write $d(x,y) = \rho(x,y)+\frac12(w(x)-w(y))$, where $\rho$ is a metric given by $\rho(x,y)=\frac12(d(x,y)+d(y,x))$ (a symmetrisation of the quasi-metric $d$). 

It can be easily shown that the asymmetry in a co-weightable quasi-metric is solely due to 
the weight function not being constant; indeed, a co-weightable quasi-metric is a metric if and only if the weight function is constant on the whole underlying set. More generally, let $\Omega(z)$ denote the set $\{x\in\Omega:w(x)=z\}$. We call $\Omega(z)$ for any $z\in w(\Omega)$ a \emph{fibre} of $\Omega$. It has been shown \cite{Vi99} that every co-weightable quasi-metric can be decomposed into a pairwise disjoint union of fibres, where each fibre is a metric space. Furthermore, every quasi-metric range query can be decomposed into a disjoint union of metric range queries on fibres. 

\begin{lemma}\label{lemma:1}
Let $X$ be a set and $d$ a co-weightable quasi-metric on $X$ with weight function $w$. Let $\rho$ be a metric symmetrising $d$ given by $\rho(x,y) = \frac12\left(d(x,y)+d(y,x)\right)$ for all $x,y\in X$. Denote by $\mathfrak{B}(x,\varepsilon)$ the set $\{y\in\Omega: d(x,y)\leq \varepsilon\}$ and by $\mathfrak{D}(x,\varepsilon)$ the set $\{y\in\Omega: \rho(x,y)\leq \varepsilon\}$. For $x\in\Omega$ and $z\in\R$, let $\delta(x,z)=\frac12(z-w(x))$. Then, for all $x\in\Omega,$ and $\varepsilon>0$
\[ \mathfrak{B}(x,\varepsilon) = \bigsqcup_{z\in w(\Omega)} \mathfrak{D}(x,\varepsilon+\delta(x,z))\big\vert\Omega(z).\]
\end{lemma}
\begin{proof}
Since $\Omega$ clearly decomposes into a disjoint union of fibres, we can write 
\begin{equation}\label{eqn:1}
\mathfrak{B}(x,\varepsilon) = \bigsqcup_{z\in w(\Omega)}\mathfrak{B}(x,\varepsilon)\big\vert \Omega(z).
\end{equation}
Now fix $z$ and suppose $y\in \Omega(z)$. Then $w(y)=z$ and hence
\begin{align*}
\rho(x,y) &= \frac12\big(d(x,y)+d(y,x)\big)\\
& = d(x,y) + \frac12\big(w(y)-w(x)\big)\\
& = d(x,y) + \frac12\big(z-w(x)\big)\\
& = d(x,y) + \delta(x,z).
\end{align*} 
Therefore, $\mathfrak{B}(x,\varepsilon)\vert \Omega(z) = \mathfrak{D}\big(x,\varepsilon+\delta(x,z)\big)\big\vert \Omega(z)$ and our result follows by (\ref{eqn:1}).
\end{proof}

The sets $\mathfrak{B}(\omega,\varepsilon)$ and $\mathfrak{D}(\omega,\varepsilon)$ defined in Lemma \ref{lemma:1} correspond, respectively, to the quasi-metric and metric queries of radius $\varepsilon$ centered at $\omega$. Therefore, metric access methods (see below) can be used to index datasets for similarity with respect to a co-weightable quasi-metric.

\subsection{Metric Trees}

The majority of published indexing schemes for similarity search apply to datasets where the dissimilarity measure is a metric (this includes the special case of access methods for vector spaces) \cite{CNBYM,HjSa03,SRF87}. In most cases we encounter a hierarchical tree index structure where each inner node of the tree is associated with a set covering a portion of the dataset and a certification function certifying if the query ball does not intersect the covering set, in which case the node is not visited and the whole branch is pruned. The data points are stored in the leaf nodes that are retrieved and sequentially scanned only if necessary. Certification functions are mostly based on distances from points; the triangle inequality ensures that no members of the dataset satisfying the query are missed. For a more general setting, we refer the reader to our theoretical exposition \cite{PeSt06}.

Metric trees are usually balanced, providing (at least in theory) scalability with respect to the number of data points. However, it is known that hierarchical indexing structures typically suffer from the `Curse of Dimensionality'. As the dimension of the dataset grows, index performance deteriorates and is often worse than the performance of an optimised sequential scan \cite{BeyerGRS99}. This effect has been linked with the so-called
phenomenon of concentration of measure on high-dimensional structures \cite{Pe00}, which is also found in many diverse areas of mathematics and physics \cite{Gr99,Le01}.

In this paper, we use two of the well known metric access methods, the M-tree \cite{CPZ97} and the mvp-tree \cite{BoOzs97}, to index short peptide fragment datasets and to compare their performance against our own indexing scheme FSIndex.

\subsubsection{M-tree}

M-tree \cite{CPZ97} is a dynamic, paged structure that can be used to index all metric datasets. It is one of the best generic indexing structures for metric spaces available, with many improvements added over the years \cite{CPZ98a,CiPa02}. A reference C++ implementation is publicly available. Memory requirements are modest as most of the indexed points are kept on hard disk and retrieved as needed, at a cost of additional I/O accesses. 

The search tree is binary and at each node a routing object (a point from the dataset) is stored together with the radius of the ball covering all the dataset points under that node. The certification function calculates the distance from the query centre to the routing object and decides whether the query intersects the covering ball. The way the routing points are chosen and data points divided between them at construction time is determined by the user's choice of one of many available split policies. Ciaccia \textit{et al.} \cite{CPZ97} found that the best performing split policies were based on generalised hyperplane decomposition where each data object is assigned to the routing object closest to it. However, such policies may be very computationally expensive (quadratic in the number of data points).

\subsubsection{mvp-tree}

The vp-tree (vantage point tree) \cite{Yianilos93} and the mvp-tree (multiple vantage point tree) \cite{BoOzs97} access methods partition the datasets using the distances from vantage points: points with a distance from a vantage point smaller than the median distance go to one branch, those with larger to the other. This procedure ensures a balanced tree. The mvp-tree is a modification of the vp-tree that uses multiple vantage points at each node of the search tree. For example, binary mvp-tree uses two instead of three vantage points needed by vp-tree to divide a covering region into four regions, resulting in fewer distance computations.

\subsection{Suffix arrays}

Trie, suffix tree and suffix array data structures form the basis of many of the established string search methods and provide the foundation for some features of the FSIndex access method described in Section \ref{sec:FSIndex}.

A \emph{trie} \cite{Fredkin60} is an ordered tree structure for storing strings having one node for every common prefix of two strings. The strings are stored in extra leaf nodes. A {PATRICIA tree} (Practical Algorithm to Retrieve Information Coded in Alphanumeric \cite{Morrison68}) is a compact representation of a trie where all nodes with one child are merged with their parent. Tries and PATRICIA trees can be used for exact matching of strings and take linear time in the length of the query string.

Now consider a single (long) string $t$ of length $m$. The \emph{suffix tree} \cite{Weiner73} for $t$ is the PATRICIA tree of the suffixes of $t$ and can be constructed in $O(m)$ time \cite{Weiner73,McCreight76,Ukkonen92}. Suffix trees, in their original form as well as generalised to suffixes of more than one string, can be used to solve a great variety of problems involving matching substrings of long strings \cite{Gusfield97}.

Suffix trees in general occupy $O(m)$ space, where the constant may be large depending on the application \cite{Gusfield97,Kurtz99}. The \emph{suffix array} data structure, first proposed by Manber and Myers \cite{Manber93}, is a compact representation of the suffix tree for $t$ consisting of the array $pos$, of integers in the range $I_m$ specifying the lexicographic ordering of suffixes of $t$, and the array $lcp$, where $lcp[i]$ contains the longest common prefix of the substrings starting at positions $pos[i-1]$ and $pos[i]$. Efficient $O(m)$ construction algorithms exist and using binary search on array $pos$ and the $lcp$ values, it is possible to search for occurrence of a string $p$ in $t$ in $O(n + \log m)$ time, where $n$ is the length of $p$ \cite{Gusfield97}. 

PATRICIA trees (and hence suffix trees and arrays), being compact representations of a set of strings, can also be used to accelerate non-exact searches \cite{Gonnet:1992}, by partially evaluating the similarity function at each inner node at pruning those branches where it exceeds the given range.

\section{FSIndex}\label{sec:FSIndex}

\emph{FSIndex} is an access method for short peptide fragment workloads mainly based on two procedures: amino acid alphabet reduction and combinatorial generation.

For very short fragments (lengths 2-4), the number of all possible fragment instances is very small (for length 3, $20^3 = 8000$) and almost every fragment instance generated exists in the dataset. Hence, it is possible to enumerate all neighbours of a given point in a very efficient and straightforward manner using digital trees or even hashing. For larger lengths, the number of fragments in a dataset is generally much smaller than the number of all possible fragments and generation of neighbours is not feasible. If it were to be attempted, most of the computation would be spent generating fragments that do not exist in the dataset. Hence the idea of mapping peptide fragment datasets to smaller, densely and, as much as possible, uniformly packed spaces where the neighbours of a query point can be efficiently generated using a combinatorial algorithm.

Partitions of the amino acid alphabet provide the means to achieve the above. Amino acids can be classified by chemical structure and function into groups such as hydrophobic, polar, acidic, basic and aromatic. Such classification appears in every undergraduate text in biochemistry and has been previously used in sequence pattern matching \cite{Smith90}. In general, substitutions between the members of the same group are more likely to be observed in closely related proteins than substitutions between amino acids of markedly different properties. The widely used similarity score matrices such as PAM \cite{Dayhoff:1978} or BLOSUM \cite{Henikoff:1992} are derived from target frequencies of substitutions and therefore capture these relationships more precisely.

The required mapping is constructed as follows. Given a set of fragments of fixed fragment length $\Sigma^m$, an alphabet partition $\pi_i:\Sigma\to\Gamma_i$ is chosen for each position $i\in I_m$, where $\abs{\Gamma_i}<\abs{\Sigma}$, canonically inducing the mapping $\pi:\Sigma^m\to\mathscr{B}$ where $\mathscr{B}=\Gamma_0\times \Gamma_1\times\ldots\times \Gamma_{m-1}$. We call the members of $\mathscr{B}$ \emph{bins} or \emph{blocks} and denote their number by $N$. An important consequence of such mapping is that the lower bounds of the value of the query function on each bin are easy to compute. 

\begin{example}\label{ex:2}
Consider the alphabet and similarity measures from Example \ref{ex:1}. A logical partition of $\Sigma$, based on the distance matrix $D$, is into sets $\alpha=\{a,c\}$ and $\beta=\{b,d\}$. Hence, for fragments of length $m=3$ with equal partitions at each position, we have $N=2^3=8$ bins.
\end{example}

It should be noted that the partitions $\pi_i$ can, but are not required to, be equal for each $i$. Different partitions at the different positions need to be allowed in order to have a finer control on the number of bins. For example, consider fragments of length $12$. If we take $5$ partitions at each position, the total number of bins would be $5^{12}\approx 244$ million while for $6$ partitions we have $6^{10}\approx 2176$ million. If the size of our dataset lies somewhere in between these numbers, using $6$ partitions would result in too many empty bins (and may be infeasible due to space requirements) while we can further improve on $5$ partitions by having $6$ partitions at only some positions.

\subsection{Data structure}

The FSIndex data structure consists of three arrays: $frag$, $bin$ and $lcp$. The array $frag$ contains pointers to each fragment in the dataset and is sorted by bin. The array $bin$, of size $N$ is indexed by the rank of each bin and contains the offset of the start of each bin in $frag$. The bin ranking function $r:\mathscr{B}\to I_N$ is defined as follows. For each $i\in I_m$ let $r_i:\Gamma_i\to I_{\abs{\Gamma_i}}$ be a ranking function of $\Gamma_i$ and define $\xi_i:\Gamma_i\to\N$ by
\begin{equation}\label{eq:FSrank1}
\xi_i(\gamma) = r_i(\gamma) \prod_{j=i}^{m-1}\abs{\Gamma_j}.
\end{equation}
In the case $i=m-1$ the empty product above is taken to be equal to $1$. Then, for any $B=B_0B_1\ldots B_{m-1} \in\mathscr{B}$,
\begin{equation}\label{eq:FSrank2}
r(B)= \sum_{i=0}^{m-1} \xi_i(B_i).
\end{equation}

\begin{figure}[!ht] 
\begin{center}
\scalebox{0.7}{\includegraphics{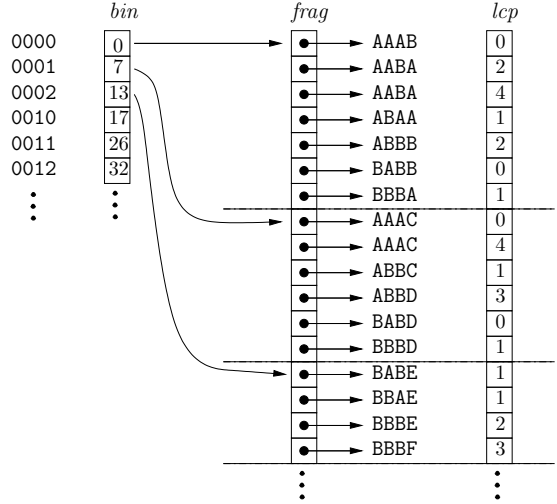}}
\caption[Structure of an FSIndex.]{Structure of an FSIndex of a dataset of fragments of length 4 from the alphabet $\Sigma=\{{\tt A,B,C,D,E,F}\}$. The same alphabet reduction is used at each position, mapping $\{{\tt A,B}\}$ to ${\tt 0}$, $\{{\tt C,D}\}$ to ${\tt 1}$ and $\{{\tt E,F}\}$ to ${\tt 2}$.}
\label{fig:FSIndexst}
\end{center}
\end{figure}

In addition, each bin is sorted in lexicographic order and the value of $lcp[i]$ provides the length of the longest common prefix between $frag[i]$ and $frag[i-1]$. The value of $lcp[0]$ is set to $0$. Figure \ref{fig:FSIndexst} depicts an example of the full structure of an FSIndex.

The $frag$ and $lcp$ arrays are very similar to suffix arrays but the order of offsets in $frag$ is different because $frag$ is first sorted by bin and then each bin is sorted in lexicographic order. Sorting $frag$ within each bin and constructing and storing the $lcp$ array is not strictly necessary and incurs a space and construction time penalty. The benefit is improved search performance for large bins, compensating for unbounded bin sizes. In effect, each bin is subindexed using a compact version of a PATRICIA tree.

\subsection{Index construction}

The construction algorithm (Algorithm \ref{alg:FSconstruct}) is closely related to counting sort \cite{Seward:1954}. It makes three passes over data fragments: to count the number of fragments in each bin, to insert the fragments into the $frag$ array and to compute the $lcp$ array.

\begin{algo}[ht!]
\begin{pseudocode}{ConstructFSIndex}{a}\label{alg:FSconstruct}
X - \text{fragment dataset (size $n$)}\\
tmp - \text{auxillary array used to keep track of bin sizes}\\
\\
\COMMENT{First pass to get size of each bin}\\
\FOR j \GETS 0 \TO N+1 \DO
  tmp[j] \GETS 0\\
\FOREACH s\in X \DO
\BEGIN
  k \GETS r(\pi(s))\\
  tmp[k+2] \GETS tmp[k] + 1\\
\END\\
\\
\COMMENT{Second pass to insert fragments into bins}\\
\FOR j \GETS 2 \TO N+1 \DO
  tmp[j] \GETS tmp[j]+tmp[j-1]\\
\FOREACH s\in X \DO
\BEGIN
  k \GETS r(\pi(s))\\
  frag[tmp[k+1]] \GETS s\\
  tmp[k+1] \GETS tmp[k+1] + 1\\
\END\\
\\
\COMMENT{Sort each bin}\\
\FOR j \GETS 0 \TO N-1 \DO
\BEGIN
  bins[j] = tmp[j]\\
  \CALL{QuickSort}{frag[tmp[j]:tmp[j+1]]}\\
\END\\
bins[N] = tmp[N]\\
\\
\COMMENT{Compute the longest common prefixes}\\
lcp[0] \GETS 0\\
\FOR i \GETS 1 \TO n-1 \DO
\BEGIN
  x \GETS frag[i-1]\\
  y \GETS frag[i]\\
  j \GETS 0\\
  \WHILE x_j = y_j \DO j \GETS j+1\\
  lcp[i] \GETS j
\END
\end{pseudocode}
\end{algo}

The fragment dataset is in practice always obtained from a full sequence dataset by iterating over all subfragments of length $m$ from each sequence and it is often necessary to verify each fragment and reject those that contain non-standard letters such as `X', `B' or `Z' that do not represent actual amino acids and violate the triangle inequality for the score matrices. Therefore, the true number of data points is not known exactly before the first pass through the dataset. 

The space requirement of FSIndex is $\Theta(n+N)$. The exact time complexity of the construction algorithm depends on the sorting algorithm used for sorting the $frag$ array. The implementation shown in Algorithm \ref{alg:FSconstruct} uses quicksort \cite{Hoare:1962} and hence the total construction time is $O(n + N + n\log n)$ on average and $O(n+N+n^2)$ in the worst case.

\subsection{Search}

Search using FSIndex is based on traversal of implicit trees whose nodes are associated with reduced fragments (bins).

Let $B=B_0B_1\ldots B_{m-1} \in\mathscr{B}$. For each $k\in I_m$ and $\gamma\in\Gamma_k$, denote by $B(k,\gamma)$ the sequence $B_0\ldots B_{k-1}\gamma B_{k+1} \ldots B_{m-1}$. For every $i\in I_m$ denote by $T_{B,i}$ the tree having the root $B$ connected to the subtrees $T_{B(k,\gamma),k+1}$ for all $k=i,i+1,\ldots, m-1$ and $\gamma\in\Gamma_k\setminus\{B_k\}$ and by $T_{B}$ the tree $T_{B,0}$.

\begin{figure}[!ht] 
\begin{center}
\scalebox{0.6}{\includegraphics{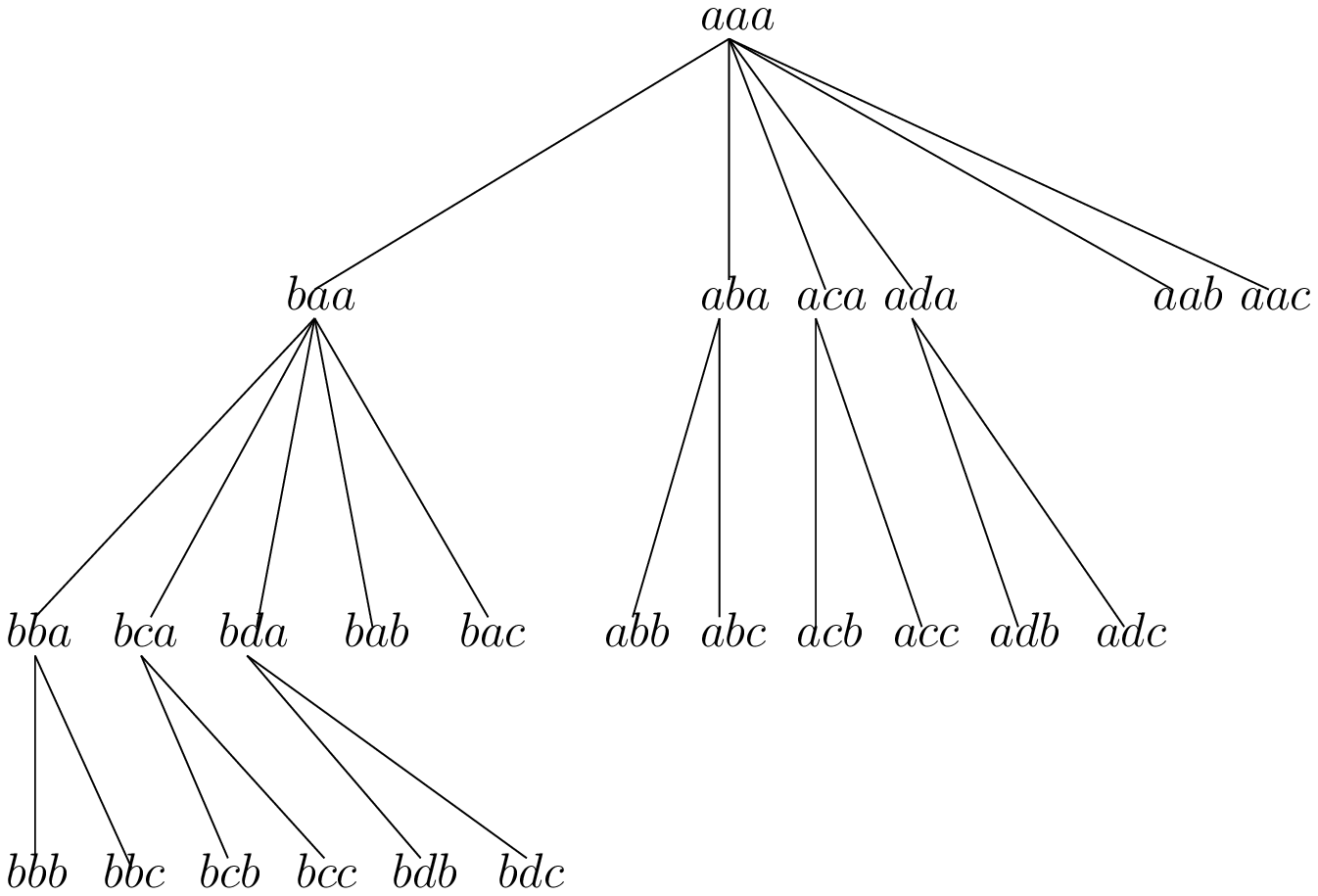}}
\caption{An example of $T_{B}$ where $B=aaa\in\Gamma_0\times\Gamma_1 \times\Gamma_2$, $\Gamma_0=\{a,b\}$, $\Gamma_1=\{a,b,c,d\}$, $\Gamma_2=\{a,b,c\}$.}
\label{fig:FStree}
\end{center}
\end{figure}

The trees $T_{B,i}$ are connected and unbalanced and can be shown to have depth $m-i$ while the root has the degree $\sum_{k=i}^{m-1}\abs{\Gamma_k}-1$. The topology of $T_B$ depends only on the sizes of the alphabets $\Gamma_i$ and not on any particular $B\in\mathscr{B}$. If $\abs{\Gamma_0}= \abs{\Gamma_1}=\ldots =\abs{\Gamma_{m-1}}=K$, $T_{B}$ is isomorphic to the \emph{multinomial tree} of order $(m,K)$. If $K=2$, such tree is called the \emph{binomial tree} of order $m$. An example is shown in Figure \ref{fig:FStree}. 

It is easily established that for any $B\in \mathscr{B}$ there exists a bijection between the nodes of $T_B$ and the set $\mathscr{B}$, that is, traversal of $T_B$ enumerates $\mathscr{B}$. It can also be easily seen that the nodes at depth $k$ from the root have Hamming distance $k$ from the root.

\subsubsection{Range queries}

Retrieval of a half-plane range query $Q^\text{rng}_f(\varepsilon)$ using the implicit tree structure is conceptually straightforward. Define $F:\mathscr{B}\to\R$ by $F(B)=\sum_{i=0}^{m-1} F_i(B_i)$ where $F_i(\gamma)= \min_{a\in \gamma} f(a)$ so that $F$ gives the lower bound for $f$ over each bin. Let $Z_i=\argmin_{\gamma \in\Gamma_i} F_i(\gamma)$ and let $Z=Z_0Z_1\ldots Z_{m-1}\in\mathscr{B}$.  We traverse the tree $T_{Z}$ from the root, calculating at each node $B$ the value of $F(B)$ and pruning the subtree rooted at $B$ if $F(B)>\varepsilon$. For every visited node that is not pruned, we sequentially scan the associated bin and collect all the fragments $x\in\Sigma^m$ such that $f(x)\leq\varepsilon$.

By the choice of $Z$, we have $F(Z)\leq F(B)$ for all $B\in\mathscr{B}$. Furthermore, since each branch in $T_Z$ involves a change of single letter at a position, it is easy see that for each child $C$ of $B$, $F(B)\leq F(C)$. This implies that if $F(B)>\varepsilon$, pruning the subtree rooted at $B$ will not lose any hits. Therefore, FSIndex provides what we call a \emph{consistent indexing scheme} \cite{PeSt06}.

Our implementation of range search first processes (sequentially scans) the root bin $Z$ and then performs a recursive, depth-first traversal of the implicit tree $T_Z$ using the function \textsc{CheckNode} (Algorithm \ref{alg:FSchkbin}) with the initial arguments $u=r(Z)$, $D=0$ and $i=0$. The values of $F_k(\gamma)$, $\min\big\{F_k(\gamma)\ |\ \gamma\in\Gamma_k \setminus\{Z_k\}\big\}$ and $\xi_k(\gamma) - \xi_k(Z_k)$ are precomputed for all $k$ and all $\gamma$ before search making each evaluation of \textsc{CheckNode} very fast. All computations involve the ranks of reduced fragments rather than the fragments themselves because the $bin$ array is indexed by rank. The function \textsc{ProcessBin} (Algorithm \ref{alg:FSprocessbin}) scans each bin associated with a node that is not pruned. It uses the $lcp$ array to reduce the number of computations of the function $f$.

\begin{algo}[ht!]
\begin{pseudocode}{CheckNode}{u,D,i}\label{alg:FSchkbin}
\FOR j \GETS m-1 \DOWNTO i \DO
\BEGIN
  \IF D + \min\big\{F_j(\gamma)\ |\ \gamma\in\Gamma_j \setminus\{Z_j\}\big\} \leq\varepsilon \THEN
  \BEGIN
    \FOREACH \gamma\in\Gamma_j \setminus\{Z_j\} \DO
    \BEGIN
      E \GETS D + F_j(\gamma)\\
      \IF E\leq\varepsilon \THEN
      \BEGIN
        v \GETS u + \xi_j(\gamma) - \xi_j(Z_j)\\
        \CALL{ProcessBin}{v}\\
        \CALL{CheckNode}{v,E,j+1}\\
      \END
    \END   
  \END 
\END
\end{pseudocode}
\end{algo}

\begin{algo}[htb!]
\begin{pseudocode}{ProcessBin}{u}\label{alg:FSprocessbin}
HL -\text{list of hits}\\
CD -\text{array of length $m+1$}\\
\\
n\GETS bin[u+1]-bin[u]\\
\IF n=0 \THEN \RETURN{} \\
CD[0]\GETS 0\\
\FOR i\GETS 0 \TO n-1 \DO
\BEGIN
  s\GETS frag[u+i]\\
  \FOR j\GETS lcp[u+i] \TO lcp[u+i+1]-1 \DO
    CD[j+1] \GETS CD[j] + f_j(s_j)\\
  \IF CD[lcp[u+i+1]]\leq\varepsilon \THEN
  \BEGIN
    \FOR j\GETS lcp[u+i+1] \TO m-1 \DO CD[j+1] \GETS CD[j] + f_j(s_j)\\
    \IF CD[m]\leq\varepsilon \THEN \CALL{InsertHit}{HL,s,CD[m]} \\
  \END
\END
\end{pseudocode}
\end{algo}

\begin{example}\label{ex:3}
Continuing with the fragment space and partitions described in Examples \ref{ex:1} and \ref{ex:2}, suppose we need to retrieve the query $Q^\text{rng}_f(7)$ where $f(v)=d(x,v)$ for $x=abd$. In this case, $f_i(\gamma)=D(x_i,\gamma)$ for $i=0,1,2$. We first compute $F_i$, as shown:
\[\begin{tabular}{|l|rrr|}
\hline
&  $F_0$ &  $F_1$ &  $F_2$ \\ \hline
$\alpha$&  0 & 8 &  8\\
$\beta$& 7 &  0 & 0\\
\hline
\end{tabular}\]
For example, $F_1(\alpha)=\min\{D(b,a),D(b,c)\}=8$. 

Clearly, $Z=\alpha\beta\beta$ (the bin containing $x$) and $F(Z)=0$ so $Z$ must be scanned. The children of $Z$ are the bins $\beta\beta\beta$, $\alpha\alpha\beta$ and $\alpha\beta\alpha$ having the values of $F$ $7$, $8$ and $8$ respectively. Hence, we prune the subtrees originating at $\alpha\alpha\beta$ and $\alpha\beta\alpha$ and scan the bin $\beta\beta\beta$. Both children of $\beta\beta\beta$ have $F$ greater than $7$ so the algorithm stops here.
\end{example}

\subsubsection{$k$NN queries}

Our $k$NN search algorithms use \emph{branch-and-bound} \cite{CPZ97,HjSa03} traversal involving initially setting the radius $\varepsilon$ to a very large number, inserting first $k$ data points encountered into the list of hits and then setting $\varepsilon$ to be the largest value of $f$ for all hits. From then on, if a point having the value of $f$ less than $\varepsilon$ is found, it replaces the hit with a previously largest value of $f$ and the new value of $\varepsilon$ is computed. Eventually, the value of $\varepsilon$ is reduced to the exact value necessary to retrieve $k$ nearest neighbours.

We implement the branch-and-bound procedure using a priority queue (heap) returning the largest value of $f$ for the list of hits. Most of the code for range search can be reused: it is only necessary to change the \textsc{InsertHit} routine used by the \textsc{ProcessBin} function.

\subsection{Arbitrary fragment lengths}

In most practical situations, fragment datasets are datasets of suffixes of full sequences. The FSIndex structure as is can be used without modifications for answering queries longer than $m$, the original length: each fragment of length $m$ is a prefix of a suffix of length $m'$ where $m'\geq m$. To search with a query of length $m'$, traverse the search tree using the first $m$ positions and sequentially scan all the bins retrieved, using all $m'$ positions to calculate the value of $f$. If $m'>m$, the few fragments of length $m$ at the end of each full sequence can be identified and ignored at the sequential scan step.

Similarly, FSIndex can be used to answer queries of length $m''$ where $m''<m$. At the construction step, insert all suffixes, including those of length less than $m$ into the index by mapping each fragment $x$ such that $\abs{x}=m''<m$, into the bin $\pi_1(x_1)\pi_2(x_2)\ldots \pi_{m''}(x_{m''})\gamma_{m''+1}\ldots \gamma_{m}$, where $\gamma_{m''+1},\ldots, \gamma_{m}$ are chosen so that $\xi_{m''+1}(\gamma_{m''+1})=\xi_{m''+2}(\gamma_{m''+2})=\ldots =\xi_{m}(\gamma_{m})=0$. 

To answer a query of length $m''$, traverse the search tree up to the depth $m''$ and sequentially scan all the bins attached to subtrees rooted at the accepted nodes using first $m''$ positions to evaluate $f$. The ranking function given by the Equations \ref{eq:FSrank1} and \ref{eq:FSrank2} ensures that the bins that are the children of a given node are adjacent in the $frag$ array.

\section{Performance of FSIndex}

This section describes the experiments on actual fragment datasets carried out to evaluate the performance of FSIndex. Three main classes of tests were conducted investigating general performance, scalability and effects of similarity measures.

Each experiment consisted of 5000 searches using randomly generated queries. The queries were produced
by first generating several random sequences and then taking a sample of non-overlapping fragments. Each  full sequence was constructed using \emph{Dirichlet mixtures} \cite{SKBHKMH96,Durbin:1998}, with the code obtained from \url{http://www.cse.ucsc.edu/research/compbio/dirichlets/}: the length of the sequence was modeled by a discretised log-normal distribution \cite{PHHC00} and the amino acids of a sequence are generated by an independent, identically distributed process with the the probabilities coming from a mixture of Dirichlet densities. According to the authors of \cite{SKBHKMH96}, using Dirichlet mixtures gives a better model of protein sequences than using a single distribution based on background frequencies of amino acids.

The main measures of performance are the number of bins and dataset fragments scanned in order to retrieve $k$ nearest neighbours. The principal reason for expressing the results in terms of the number of nearest neighbours retrieved rather than the radius is that it allows comparison across different indexing schemes, datasets and similarity measures.  Furthermore, most existing protein datasets are strongly non-homogeneous and the number of points scanned in order to retrieve a range query for a fixed radius varies greatly compared to the number of points scanned in order to retrieve a fixed number of nearest neighbours. However, it should be noted that most experiments involve range search algorithms, which are generally more efficient.

The final statistic is the percentage of residues (letters) scanned out of the total number of residues in all scanned fragments. It measures the effect of sub-indexing each bin using the suffix-array-like structure involving partially scanning each fragment with a help of the $lcp$ array.

FSIndex creation and search algorithms were implemented in the C programming language.
The experiments described in \ref{sec:FSperf}, \ref{sec:FSscale}, \ref{subsec:FSperf0} and \ref{subsec:rngkNN} were were performed on a Sun Fire[tm] 280R server (733 MHz CPU, 2 GB of memory) under Solaris. The experiments described in \ref{subsec:FSsim} and \ref{subsubsec:SAMVPT} were performed on an Intel(R) XEON [tm] 2.2 GHz CPU under Linux. Performance of M-tree (\ref{subsubsec:Mtree}) was tested on Sun Fire[tm] 6800 servers (1.2 GHz CPU) under Solaris. All testing programs were compiled with the native compiler (Sun compiler on Solaris, gcc on Linux) with maximum optimisation flags.

\subsection{Datasets and indexes}

We used two publicly available protein sequence datasets as sources for our overlapping fragment datasets: NCBI nr (non-redundant -- version from June 2004 containing 1,866,121 sequences consisting of 619,474,291 amino acids) \cite{WheelerNCBI04} and SwissProt (Release 43.2 of April 2004, containing 144,731 sequences consisting of 53,363,726 amino acid residues) \cite{Boeckmann2003}. 

The experiments investigating general performance and effect of different similarity measures used fragment datasets derived from SwissProt. The scalability experiments used, in addition to SwissProt, the datasets \texttt{nr018K}, \texttt{nr036K}, \texttt{nr072K}, and \texttt{nr288K}, obtained by randomly sampling 18, 36, 72 and 288 thousands of sequences respectively from the nr dataset (SwissProt fills the gap because it contains about 150,000 sequences).  Table \ref{tbl:FSinddata} describes the datasets and alphabet partitions used for the experiments.

\begin{table*}[!ht]
\begin{center}
{\scriptsize\tt
\begin{tabular}{|l|l|l|r|r|r|}
\hline
Index & Dataset & Partitions & Fragments & Bins & \parbox{1.7cm}{\centering Space usage (Mb)} \\
\hline\hline
SPEQ06 & SwissProt & T,SA,N,ILV,M,KR,DE,Q,WF,Y,H,G,P,C & 53486349 &  7529536 & 283 \\ \hline
SPEQ09 & SwissProt & TSAN,ILVM,KR,DEQ,WFYH,GPC & 53478888 & 10077696 & 293 \\ \hline
SPEQ12 & SwissProt & TSAN,ILVM,KRDEQ,WFYHGPC & 53472161 & 16777216 & 318 \\ \hline
\hline
nr01809 & nr018K & TSAN,ILVM,KR,DEQ,WFYH,GPC & 6005750 & 10077696 & 67 \\ \hline
nr03609 & nr036K & TSAN,ILVM,KR,DEQ,WFYH,GPC & 11911191 & 10077696 & 95 \\ \hline
nr07209 & nr072K & TSAN,ILVM,KR,DEQ,WFYH,GPC & 23878523 & 10077696 & 152 \\ \hline
nr28809 & nr288K & TSAN,ILVM,KR,DEQ,WFYH,GPC & 95593618 & 10077696 & 494 \\ \hline
\hline 
SPNA09 & SwissProt & KR,Q,E,D,N,T,SA,G,H,W,Y,F,P,C,ILV,M & 53478888 & 10483200 & 294 \\
  & & KR,Q,ED,N,T,SA,G,HW,YF,P,C,ILV,M & & &\\
  & & KR,QED,N,TSA,G,HW,YF,P,C,ILVM & & &\\
  & & KR,QEDN,TSA,G,HWYF,PC,ILVM & & &\\
  & & KR,QEDN,TSA,G,HWYFPC,ILVM & & &\\
  & & KR,QEDN,TSAG,HWYFPC,ILVM & & &\\
  & & KRQEDN,TSAG,HWYFPC,ILVM & & &\\
  & & KRQEDN,TSAG,HWYFPCILVM & & &\\
  & & KRQEDNTSAG,HWYFPCILVM & & &\\ \hline
SPNB09 & SwissProt & KR,QEDN,TSA,G,HWYF,PC,ILVM & 53476582 & 8643600 & 287 \\
  & &  KR,QEDN,TSA,G,HWYF,PC,ILVM & & &\\
  & &  KR,QEDN,TSA,G,HWYF,PC,ILVM & & &\\
  & &  KR,QEDN,TSA,G,HWYF,PC,ILVM & & &\\
  & &  KR,QEDN,TSA,G,HWYFPC,ILVM & & &\\
  & &  KR,QEDN,TSAG,HWYFPC,ILVM & & &\\
  & &  KR,QEDN,TSAG,HWYFPC,ILVM & & &\\
  & &  KRQEDN,TSAG,HWYFPC,ILVM & & &\\
  & &  KRQEDN,TSAG,HWYFPCILVM & & &\\
  & &  KRQEDNTSAG,HWYFPCILVM & & &\\ \hline
\hline
\end{tabular}
}
\end{center}
\caption{Instances of FSIndex used in experimental evaluations. The last two digits of the index name denote the length of reduced fragments. The indexes \texttt{SPNA09} and \texttt{SPNB09} use non-equal partitions at different positions (all shown) while the remainder were constructed using one partition for all positions (only one shown). Space usage denotes the space used by index only, excluding the dataset itself.}\label{tbl:FSinddata}
\end{table*}

In most cases, the partitions of the amino acid alphabet were chosen as a result of practical considerations based on the BLOSUM62 quasi-metric (Figure \ref{fig:blosum62qd}) distances. The main requirement was that the distances from amino acids to clusters be as large as possible, ensuring a high lower bound on distances from any possible query point to any partition but its own. While this can be made into a combinatorial optimization problem (for example, by maximizing the expected distance from an amino acid to a cluster based on the relative frequencies of amino acids), the number of possible partitions of a set of 20 elements is extremely large and the exact solution of the problem is infeasible. Instead, the partition for the index \texttt{SPEQ06} was produced by clustering amino acids that were close according to the BLOSUM62 quasi-metric and the partitions for the indexes \texttt{SPEQ09} and \texttt{SPEQ12} were obtained by coarsening it by merging `close' clusters.

The additional requirement we considered was to have as few empty bins as possible. We attempted to keep the sizes of clusters in terms of their relative frequencies close to balanced and the number of partitions per residue was decreased with fragment length. For that reason, some amino acids having very small overall frequencies, such as tryptophan (`W') and cysteine (`C'), had often to be clustered with amino acids that were very distant from them. 

\begin{figure}[ht!]
\begin{center}
\scalebox{0.6}{\includegraphics{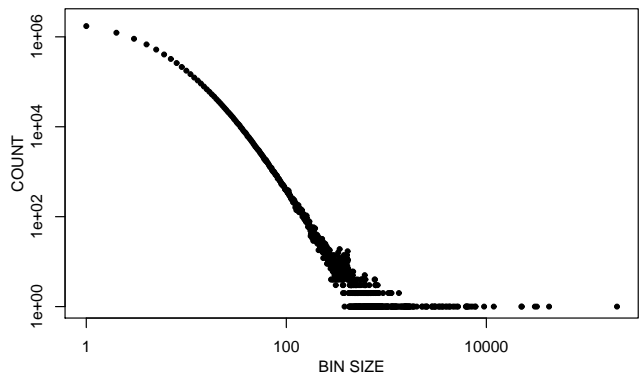}}
\caption[Distribution of \texttt{SPEQ09} bin sizes.]{Distribution of \texttt{SPEQ09} bin sizes (2,342,940 empty bins out of 10,077,696).}\label{fig:binsizedist}
\end{center}
\end{figure}

The alphabet partitions from Table \ref{tbl:FSinddata} agree with the `biochemical intuition', that is, with the chemical properties of amino acids. For example, the clusters outlined in Figure \ref{fig:blosum62qd} used for fragments of length 9 approximately correspond to polar uncharged, hydrophobic, basic, acidic, aromatic and `other' amino acids. The partition used for the fragments of length 12 is obtained by merging together acidic and basic as well as aromatic and `other' clusters. An interesting fact is that in this case each of the the four clusters has a relative frequency very close to $\frac14$. This however, does not guarantee balanced bin sizes: in all cases the distributions of bin sizes were strongly skewed in favour of small sizes (Figure \ref{fig:binsizedist} shows one example) with many empty but also a few very large bins. Such distributions appear to follow the DGX distribution, a generalisation of Zipf-Mandelbrot law proposed by Bi, Faloutsos and Korn \cite{BFK01}. 

As can be seen from Table \ref{tbl:FSinddata}, the number of bins in most cases is close to $10$ million, giving on average, in the case of the \texttt{SwissProt} dataset, about $5.3$ points per bin. The indexes \texttt{SPEQ06} and \texttt{SPEQ12} have somewhat fewer and more bins, respectively, because a closer number could not be reached with the same partitions at the same position. It was practically infeasible to test the basic case where each amino acid is given its own partition because of the large memory requirement of FSIndex in this case. Even for length 6, this would result in $20^6=64$ million bins, many of which would be empty. The large number of bins also means that storing the index on disk would result in too many page accesses and excessive runtimes.

The indexes \texttt{SPNA09} and \texttt{SPNB09} used position specific partitions in order to test the performance of unequal partitions. The choice was mainly based on biochemical categorisations, merging `close' partitions if needed. The index \texttt{SPNA09} used larger numbers of partitions at the first few positions while the numbers of partitions for \texttt{SPNB09} was more balanced.

It should be noted that it was indeed necessary to use alphabet clusters approximately corresponding to their biological characterisations (and hence to most similarity measures used in practice). Our preliminary investigations using random partitions showed that a search in that case is essentially reduced to sequential scan -- too expensive to be used for comparison. 

\subsection{General performance}\label{sec:FSperf}

\begin{figure*}
\begin{center}
\begin{tabular}[t!]{lcr}
\multicolumn{1}{l}{\mbox{\bf (a)}} &
        \multicolumn{1}{l}{\mbox{\bf (b)}} &
               \multicolumn{1}{l}{\mbox{\bf (c)}}  \\ 
\scalebox{0.79}[0.79]{\includegraphics{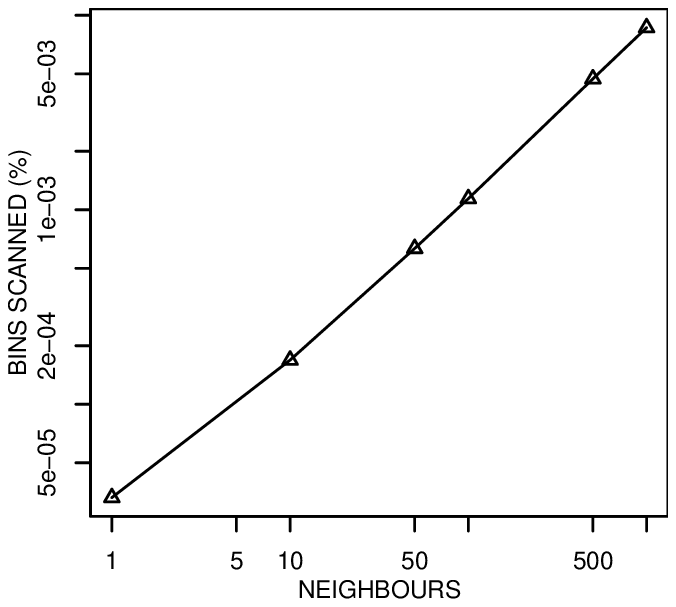}} &
\scalebox{0.79}[0.79]{\includegraphics{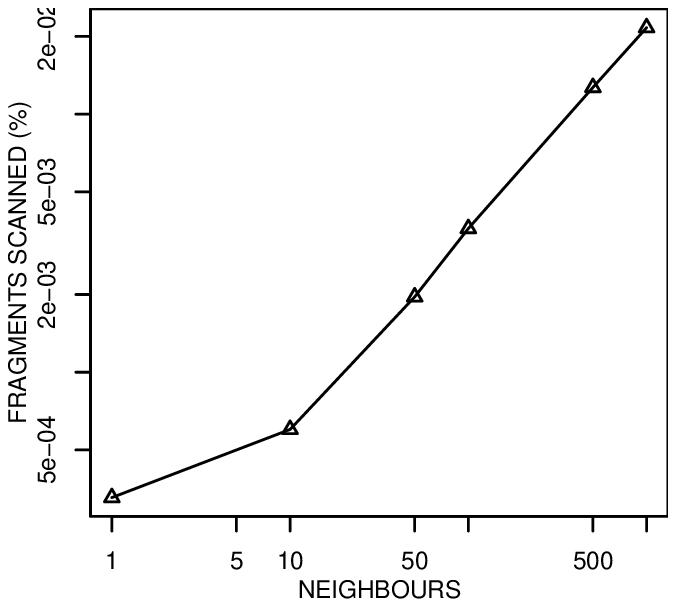}} &
\scalebox{0.79}[0.79]{\includegraphics{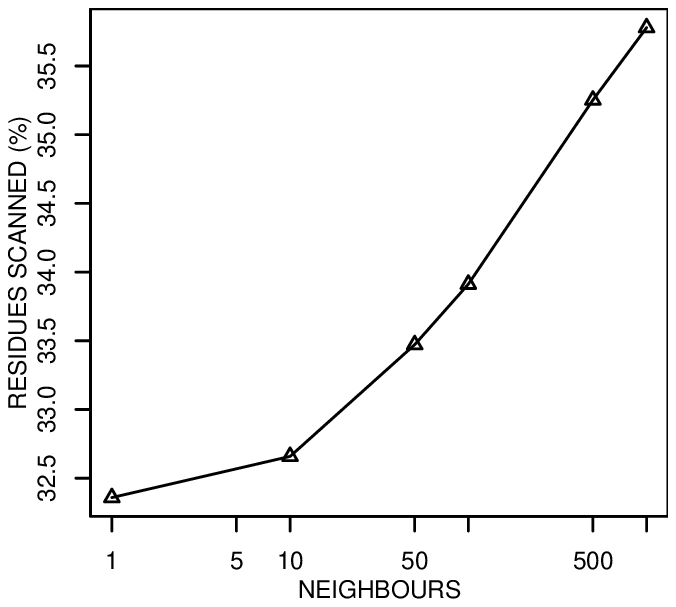}} \\ 

\multicolumn{1}{l}{\mbox{\bf (d)}} &
        \multicolumn{1}{l}{\mbox{\bf (e)}} &
               \multicolumn{1}{l}{\mbox{\bf (f)}}  \\ 
\scalebox{0.79}[0.79]{\includegraphics{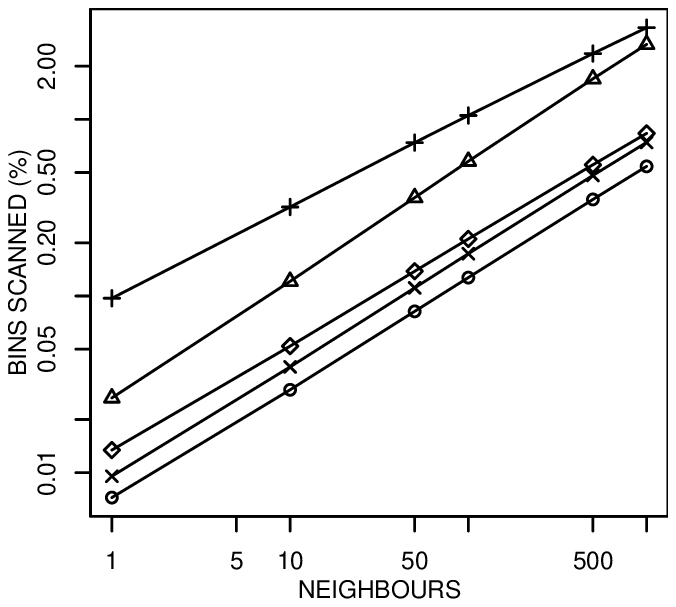}} &
\scalebox{0.79}[0.79]{\includegraphics{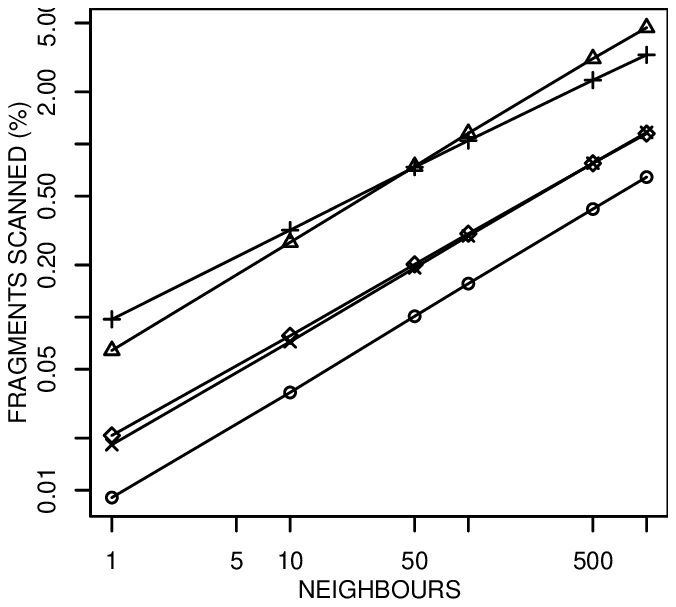}} &
\scalebox{0.79}[0.79]{\includegraphics{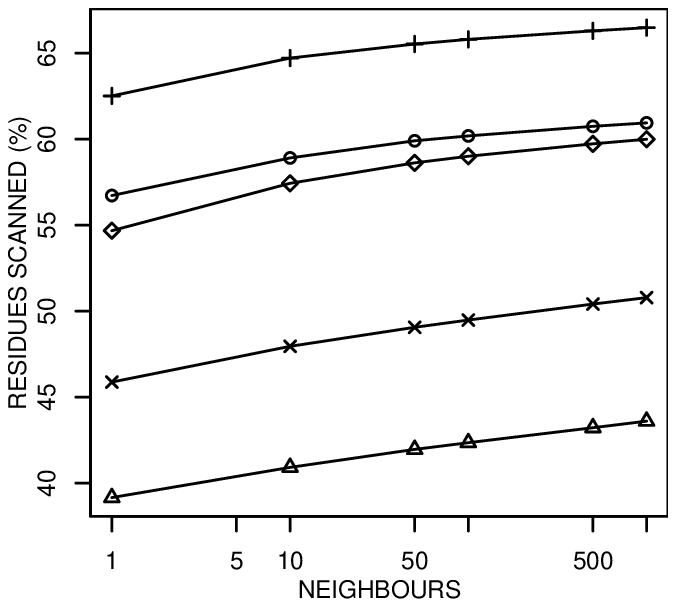}} \\ 

\multicolumn{1}{l}{\mbox{\bf (g)}} &
        \multicolumn{1}{l}{\mbox{\bf (h)}} &
               \multicolumn{1}{l}{\mbox{\bf (i)}}  \\ 
\scalebox{0.79}[0.79]{\includegraphics{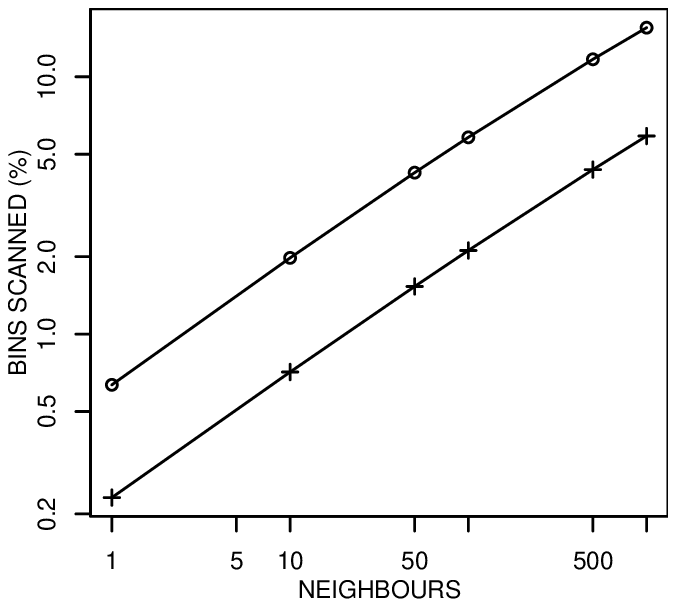}} &
\scalebox{0.79}[0.79]{\includegraphics{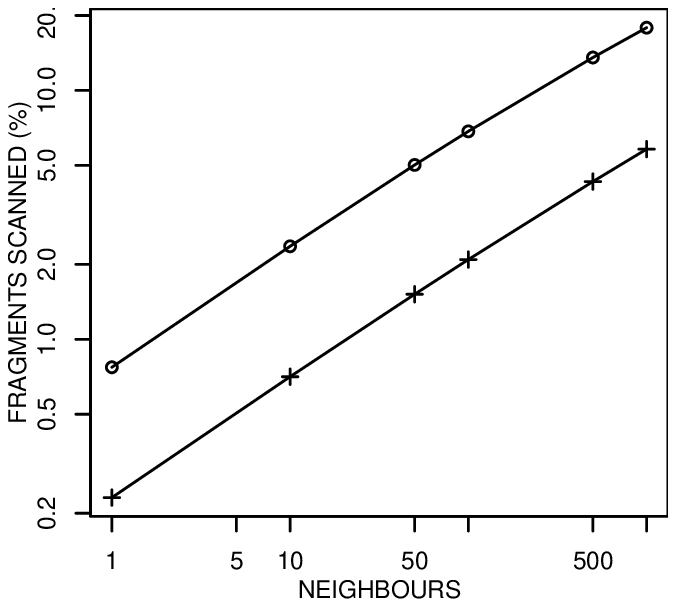}} &
\scalebox{0.79}[0.79]{\includegraphics{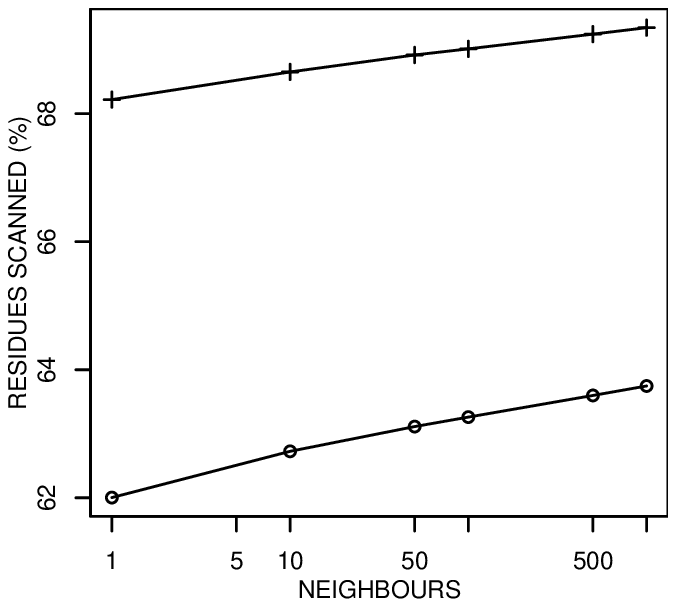}} \\ 

\end{tabular}
\begin{tabular}[c]{lr}
\mbox{
\begin{minipage}[c]{0.84\textwidth}
\caption[]{Performance of FSIndex for fragments of length 6 ({\bf a-c}), 9 ({\bf d-f}) and 12 ({\bf g-i}). First column ({\bf a,d,g}): average number of bins scanned. Second column ({\bf b,e,h}): average number of fragments scanned. Third column ({\bf c,f,i}): percentage of residues scanned (out of total number of residues in fragments scanned).
}\label{fig:FSperf}
\end{minipage}
} 
&
\mbox{
\begin{minipage}[c]{0.16\textwidth}
\scalebox{1.0}[1.0]{\includegraphics{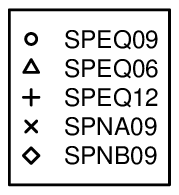}}\\
\end{minipage}
} \\
\end{tabular}
\end{center}
\end{figure*}

\begin{figure*}
\begin{center}
\begin{tabular}[t]{lcr}
\multicolumn{1}{l}{\mbox{\bf (a)}} &
        \multicolumn{1}{l}{\mbox{\bf (b)}} &
               \multicolumn{1}{l}{\mbox{\bf (c)}}  \\ 
\scalebox{0.79}[0.79]{\includegraphics{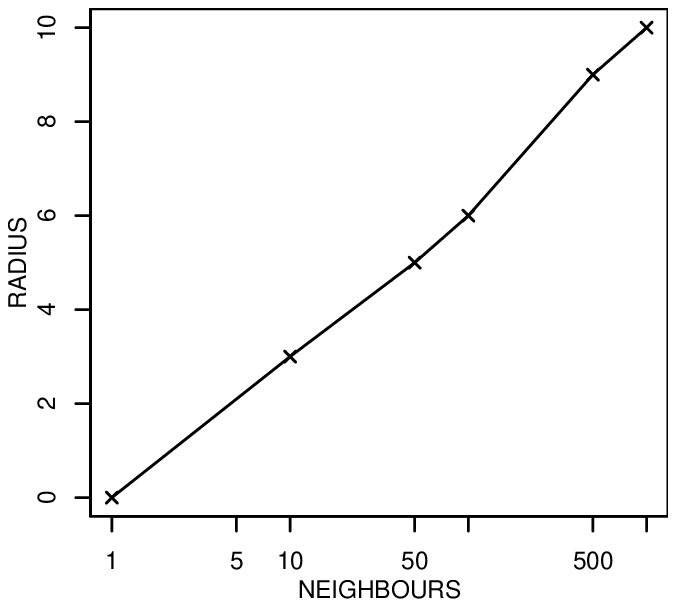}} &
\scalebox{0.79}[0.79]{\includegraphics{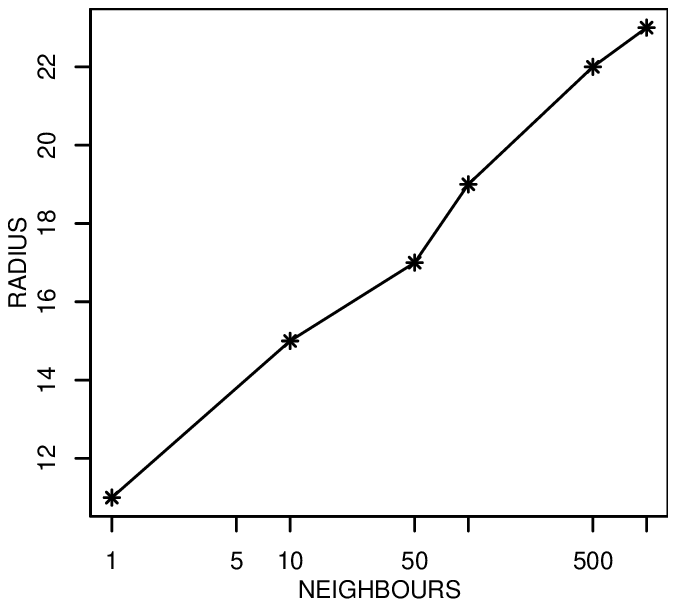}} &
\scalebox{0.79}[0.79]{\includegraphics{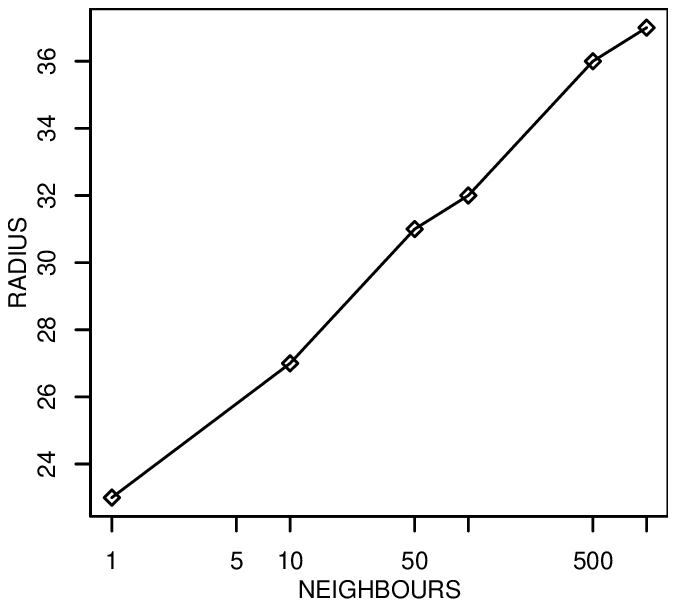}} \\ 
\end{tabular}
\caption[]{Median radius of a query retrieving $k$ nearest neighbours from the SwissProt fragment dataset of length 6 ({\bf a}), 9 ({\bf b}) and 12 ({\bf c}) with respect to the BLOSUM62 quasi-metric. 
}\label{fig:radius}
\end{center}
\end{figure*}

Figure \ref{fig:FSperf} presents selected statistics of search experiments for fragment lengths 6,9 and 12 respectively, consisting in each case of range queries retrieving 1, 10, 50, 100, 500 and 1000 nearest neighbours with respect to the BLOSUM62-based quasi-metric. For each length, $k$NN searches were performed prior to range searches using the index that was expected to be the fastest in order to determine the search ranges for each random query fragment. The median query radii required to retrieve $k$ nearest neighbours are shown in Figure \ref{fig:radius}.

\subsection{Scalability}\label{sec:FSscale}

\begin{figure*}
\begin{center}
\begin{tabular}[H]{lcr}
\multicolumn{1}{l}{\mbox{\bf (a)}} &
        \multicolumn{1}{l}{\mbox{\bf (b)}} &
               \multicolumn{1}{l}{\mbox{\bf (c)}}  \\ 
\scalebox{0.79}[0.79]{\includegraphics{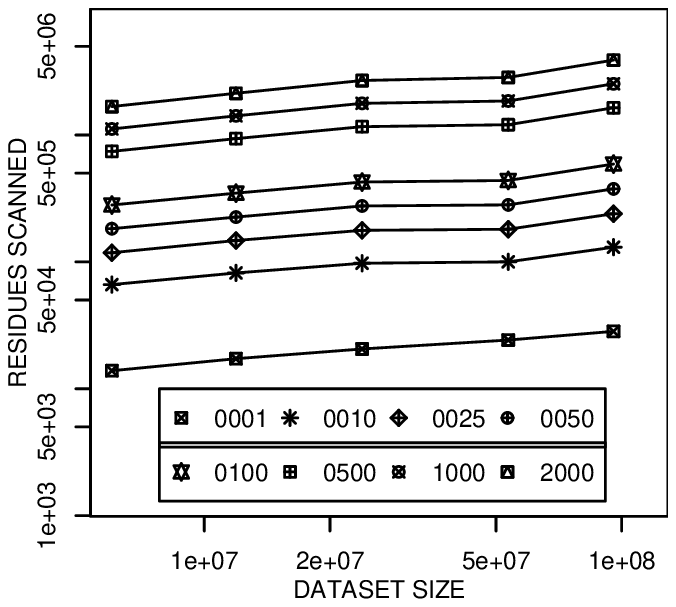}} &
\scalebox{0.79}[0.79]{\includegraphics{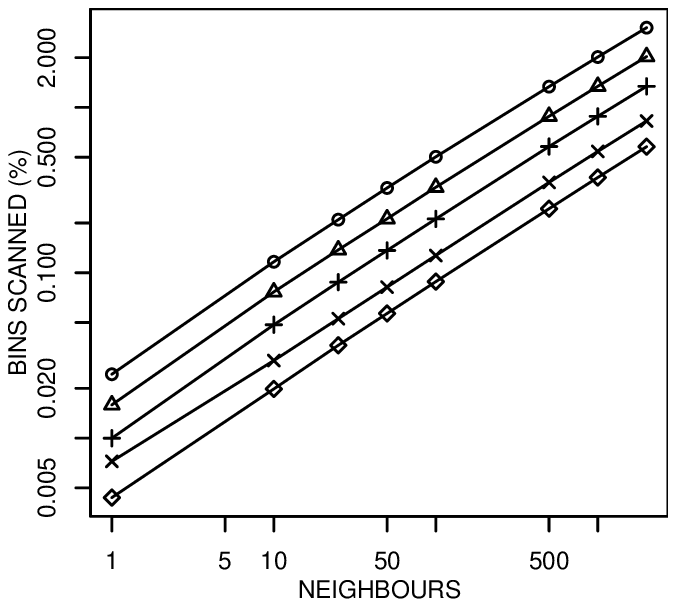}} &
\scalebox{0.79}[0.79]{\includegraphics{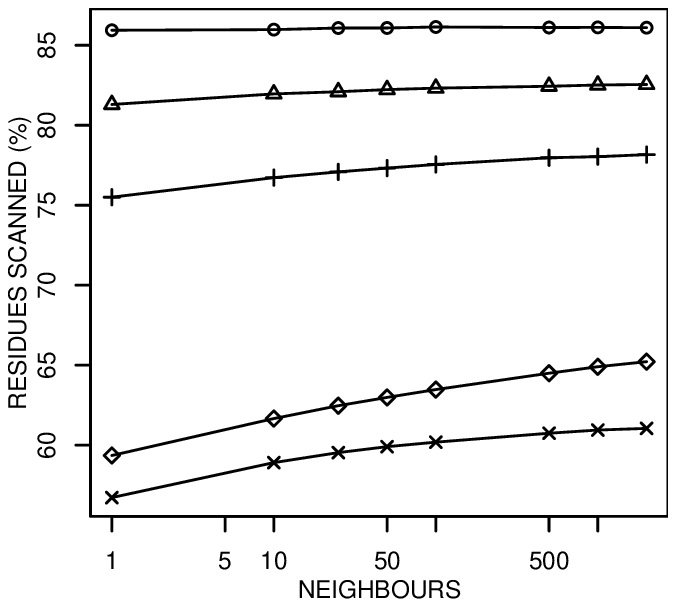}} \\ 
\end{tabular}

\begin{tabular}[H]{ll}
\mbox{
\begin{minipage}[c]{0.84\textwidth}
\vspace{-4mm}
\caption[Performance of FSIndex for fragments of length 9 (datasets of different sizes).]{Performance of FSIndex for fragment datasets of length 9 of different sizes: {\bf(a)} Scalability. Each line depicts a different number of nearest neighbours; {\bf (b)} Mean number of bins scanned; {\bf (c)} Percentage of residues scanned (out of total number of residues in fragments scanned).} \label{fig:FSscale}
\end{minipage}
} 
&
\mbox{
\begin{minipage}[c]{0.16\textwidth}
\scalebox{1.0}[1.0]{\includegraphics{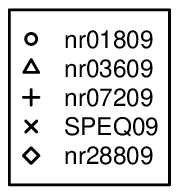}}\\
\end{minipage}
} \\
\end{tabular}
\end{center}
\end{figure*}

Figure \ref{fig:FSscale} shows the results of a set of experiments involving instances of FSIndex based on datasets of fragments of length 9 of different sizes (\texttt{nr018K}, \texttt{nr036K}, \texttt{nr072K}, \texttt{SwissProt} and \texttt{nr288K}). All indexes used the same alphabet partition (Table \ref{tbl:FSinddata}) and all queries were based on the BLOSUM62-based quasi-metric. 

\subsection{Access overhead and running time}\label{subsec:FSperf0}

\begin{figure*}
\begin{center}
\scalebox{0.7}{\includegraphics{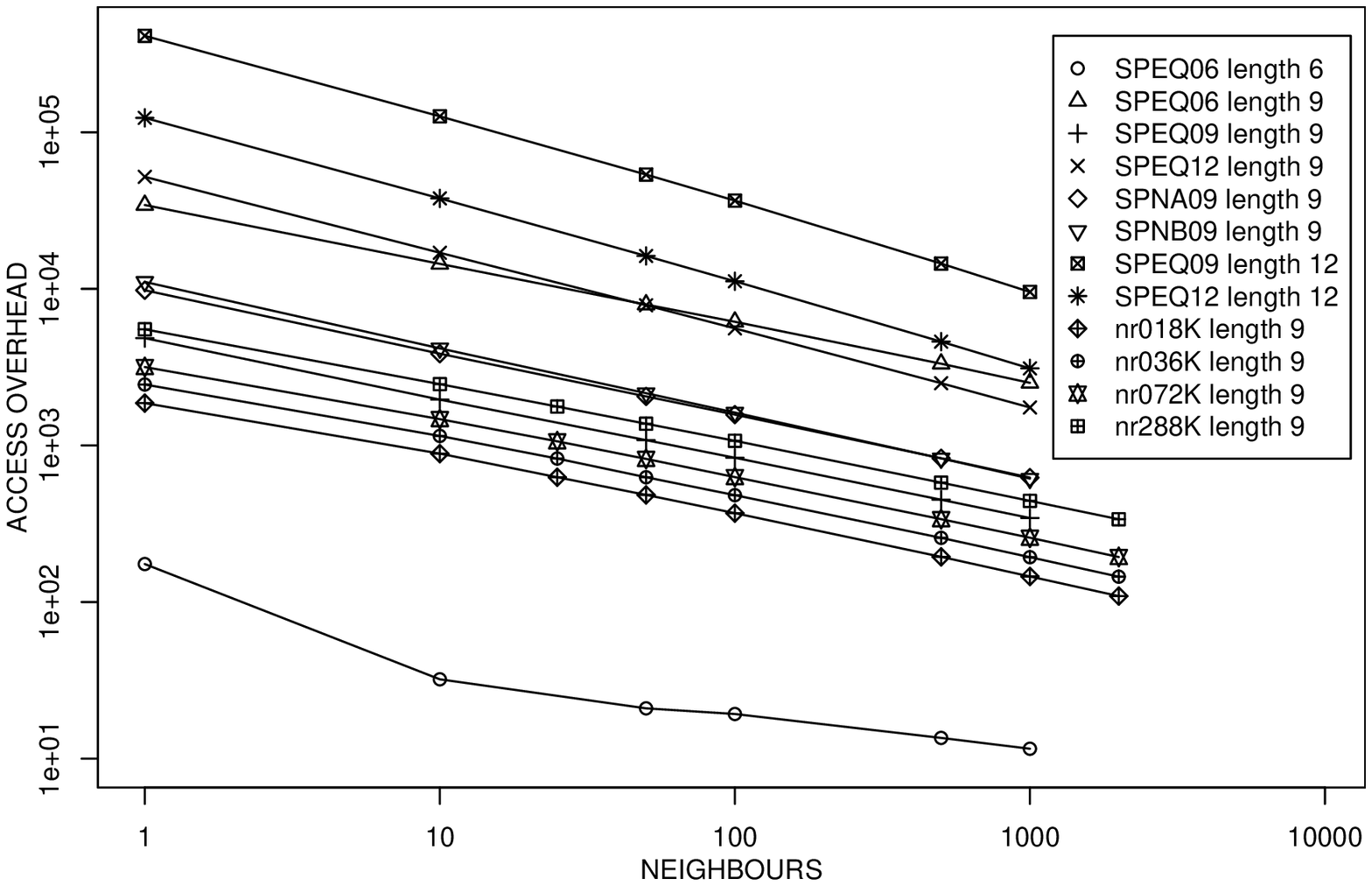}}
\caption{Average access overhead of searches using FSIndex.}\label{fig:FSaccess}
\end{center}
\end{figure*}

\begin{figure*}
\begin{center}
\scalebox{0.7}{\includegraphics{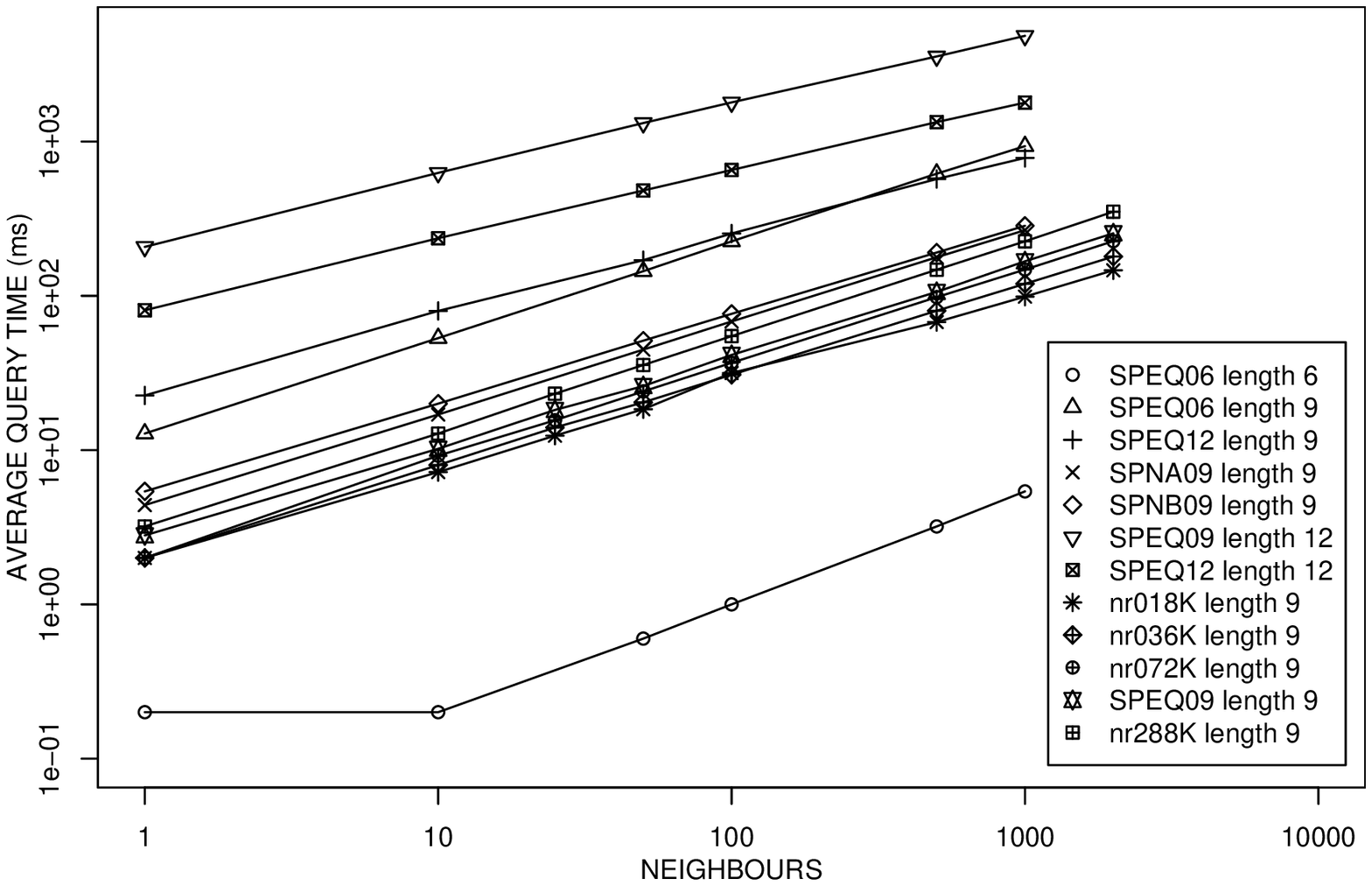}}
\caption{Average running time of searches using FSIndex.}\label{fig:FStimings}
\end{center}
\end{figure*}

Figures \ref{fig:FSaccess} and \ref{fig:FStimings} give a summary of the results of Subsections \ref{sec:FSperf} and \ref{sec:FSscale} for all combinations of indexes and fragment lengths available. Figure \ref{fig:FSaccess} shows the average access overhead, that is, the average ratio between the number of fragments scanned and the number of true neighbours retrieved. Figure \ref{fig:FStimings} shows the average search times. Range search algorithm and the BLOSUM62-based quasi-metric were used in all cases.

\subsection{Range versus $k$NN searches}\label{subsec:rngkNN}

\begin{figure}
\begin{center}
\scalebox{0.6}{\includegraphics{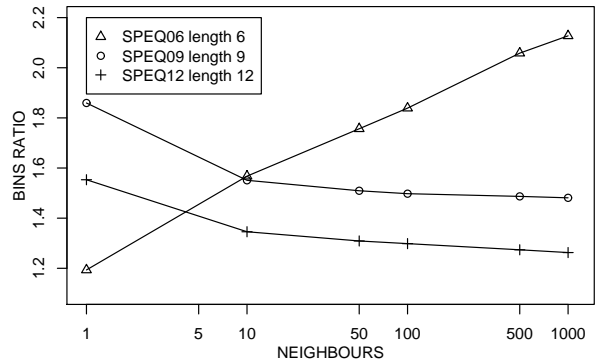}}
\caption{Mean ratio between the number of bins retrieved for $k$NN and range searches using best-performing indexes for fragment lengths $6$, $9$ and $12$.}\label{fig:FSkNN}
\end{center}
\end{figure}

Figure \ref{fig:FSkNN} shows the relative overhead of the branch-and-bound $k$NN search algorithm as compared to the range search algorithm. It gives the ratio between the number of bins retrieved for $k$NN and range searches for the best-performing indexes for each of fragment lengths $6$, $9$ and $12$.

\subsection{Dependence on similarity measures}\label{subsec:FSsim}

To investigate the difference in performance for different BLOSUM matrices, we ran range queries retrieving 100 nearest neighbours of query fragments of length 9 using the best-performing index for length 9 (\texttt{SPEQ09}). We also performed searches using the PSSMs constructed for each query fragment from the results of a BLOSUM62-based 100 NN search in order to gain some impression about usability of FSIndex for queries involving PSSMs that are likely to arise during biological applications. Since our PSSM construction routines were written in the Python programming language, we used a Python program for PSSM searches, with the FSIndex C routines embedded. Table \ref{tbl:FSmatdata} presents a summary of the results.

\begin{table}[!ht]
\begin{center}
{\small
\begin{tabular}{|l|r|r|r|r|r|}
\hline
Score Matrix & Bins (\%) & Fragments (\%) & Residues (\%) & $k$NN Ratio & Average time (ms) \\
\hline\hline
BLOSUM45 & 0.13 & 0.15 & 56.9 & 1.51 & 20 \\
BLOSUM50 & 0.13 & 0.14 & 57.1 & 1.50 & 19 \\
BLOSUM62 & 0.13 & 0.15 & 57.0 & 1.50 & 20 \\
BLOSUM80 & 0.14 & 0.16 & 57.2 & 1.48 & 21 \\
BLOSUM90 & 0.15 & 0.19 & 57.2 & 1.48 & 25 \\
PSSM & 0.11 & 0.12 & 52.0 & 9.95 & 16 \\

\hline
\end{tabular}
}
\end{center}
\caption{Performance of the FSIndex \texttt{SPEQ09} with different similarity measures. The values shown are averages based on 100 NN queries of length 9. The columns denote the similarity measure (matrix), percentages of bins, fragments and residues (as before the percentage is out of the total number of residues in scanned fragments) scanned, the ratio between the number of bins retrieved for $k$NN and range searches and the average search time.} \label{tbl:FSmatdata}
\end{table}

\subsection{Comparisons with other access methods}\label{subsec:FScomp}

The final set of experiments compares FSIndex with M-tree, mvp-tree and a suffix array based access method. In general, these access methods take significantly more space and time compared to FSIndex and it was therefore necessary to restrict the comparisons to small datasets and queries retrieving fewer neighbours.

\subsubsection{M-tree}\label{subsubsec:Mtree}

Recall that M-tree is a paged metric access method that stores the majority of the structure in secondary memory, usually on hard disk. This is in contrast with the implementations of FSIndex, mvp-tree and suffix arrays used here, which store the whole index structure in primary memory. Hence, although M-tree occupies large amounts of space, most of the costs are associated with the secondary memory, which is much less expensive in terms of price. On the other hand, I/O costs, not considered here, can be quite large.

M-tree was tested as a part of the FMTree indexing scheme that allows use of metric indexing schemes for retrieval of quasi-metric queries. FMTree was constructed by splitting the dataset into fibres and indexing each fibre separately using an instance of M-tree that was created using the BulkLoading algorithm of Ciaccia and Patella \cite{CP98}. The implementation of M-tree was obtained from its authors' web page \url{http://www-db.deis.unibo.it/Mtree/index.html}.

\begin{figure}[!htb]
\centerline{\scalebox{0.55}{\includegraphics{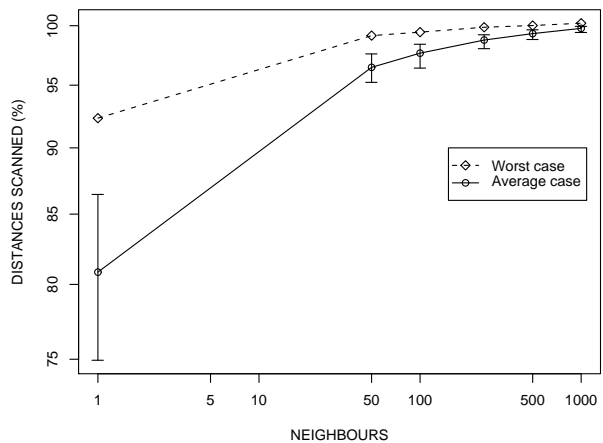}}}
\caption[Performance of FMTree based on M-tree on a dataset of fragments of length 10.]{Performance of FMTree based on M-tree on a dataset of fragments of length 10. Median and worst case results for 100 random queries are shown. Error bars show the interquartile range.}\label{fig:FMTNNperf}
\end{figure}

The dataset in this experiment was the set of 1,753,832 unique fragments of length 10 obtained from a 5000 protein sequence random sample taken from SwissProt (Release 41.21). An FMTree was generated for BLOSUM62-based quasi-metric at a cost of 34,142,940 distance computations. Figure \ref{fig:FMTNNperf} shows the results based on 100 random queries.

\subsubsection{Suffix arrays and mvp-tree}\label{subsubsec:SAMVPT}

Table \ref{tbl:FScomp} presents the results of comparisons between FSIndex ($k$NN and range search algorithm), suffix array and mvp-tree over the datasets of fragments of length 6 and 9 from \texttt{nr018K}. The similarity measure used was the associated metric to the BLOSUM62-based quasi-metric given for each $x,y\in\Sigma^m$ by $\rho(x,y)=\max\{d(x,y),d(y,x)\}$. We used a metric rather than a quasi-metric because construction of an FMtree based on mvp-tree would require significant modifications to the original code and we observed that the performance of FSIndex does not much differ if a quasi-metric is replaced by its associated metric. If the mvp-tree showed good performance on metric workloads, the next step would be to split the datasets into fibres to create an FMTree for quasi-metric searches.

Instances of suffix array were constructed using the code from \url{http://www.cs.dartmouth.edu/~doug/sarray/}. The search algorithm was identical to the Algorithm \ref{alg:FSprocessbin} where the input is a single bin containing all fragments in the dataset. In order to construct an instance of mvp-tree, duplicate fragments in the datasets were collected together and the sets of unique fragments provided to the mvp-tree construction algorithm. We modified the mvp-tree implementation developed by the original authors of mvp-tree \cite{BoOzs97} provided to us by Marco Patella for use with protein fragments. The maximum size of a leaf node was set to be 5.

\begin{table}[!ht]
\begin{center}
{\small
\begin{tabular}{|c|r||r|r||r|r||r|r||r|r|}
\hline
Length & NN & \multicolumn{2}{c||}{FSIndex ($k$NN)} & \multicolumn{2}{c||}{FSIndex (range)} & \multicolumn{2}{c||}{Suffix array} & \multicolumn{2}{c|}{mvp-tree} \\ 
 & & Overhead & Time & Overhead & Time & Overhead & Time & Overhead & Time \\ \hline\hline
6 &  1 &       15.0 &     0.1 &        9.9 &     0.1 &    20395.1 &    62.8 &     8377.6 &     4.8 \\ \hline
6 & 10 &       12.1 &     0.2 &        7.1 &     0.1 &     3809.0 &    84.1 &     6614.3 &    38.8 \\ \hline
9 &  1 &     1869.7 &     1.6 &     1303.6 &     1.3 &    73374.7 &   135.9 &  1021975.1 &   689.0 \\ \hline
9 & 10 &      902.7 &     4.5 &      615.5 &     3.3 &    14975.5 &   188.5 &   214246.4 &  1491.4 \\ \hline
%
\end{tabular}
}
\end{center}
\caption{Comparison of performance of FSIndex, suffix array and mvp-tree. The table shows the values of the average effective access overhead and the average running time for each access method. The overhead measures the number of characters (residues) accessed in order to retrieve a given number of nearest neighbours, normalised by the fragment length and the number of retrieved neighbours. The statistics are in terms of characters rather than data points because suffix array search algorithm passes by each point but only computes the distances if necessary.}\label{tbl:FScomp}
\end{table}

\section{Discussion}

While the experiments presented in the previous section covered few datasets and a small proportion of possible parameters for FSIndex creation, it can still be observed that FSIndex performed well according to many theoretical benchmarks as well as the actual search times. Not only did it perform much better than the other indexing schemes tested but it has proven itself to be very usable in practice: it does not take too much space (5 bytes per residue in the original sequence dataset plus a fixed overhead of the $bin$ array), considerably accelerates common similarity queries and the same index can be used for multiple similarity measures without significant loss of performance. In the remainder of the current section we will examine some salient features of the experimental results.

\subsection{Performance as power law}

The most striking feature of Figures \ref{fig:FSperf}-\ref{fig:FStimings} is the apparent power-law dependence of the number of bins scanned and number of fragments scanned on the number of actual neighbours retrieved, manifesting as straight lines on the corresponding graphs on the log-log scale. For each index, the slopes of the graphs of running time, bins scanned and fragments scanned are very close, implying that the same power law governs the dependence of all three variables on the number of neighbours retrieved. The exponents are 0.81 for length 6, between 0.57 and 0.63 for length 9, and about 0.45 for length 12. While a rigorous theory, especially in the context of quasi-metrics, is still missing, we believe that the exponent is related to the `dimensionality' of the index. For more detailed explanation we refer the reader to our theoretical exposition on indexing schemes \cite{PeSt06}.

\subsection{Effect of subindexing of bins}

PATRICIA-like subindexing of bins was introduced in order to accelerate scanning of bins containing many duplicate or highly similar fragments. Figures \ref{fig:FSperf} (c,f,i) and \ref{fig:FSscale} (c) show that there are two main factors influencing the proportion of residues scanned out of the total number of residues in the fragments belonging to the bins needed to be scanned: the (average) size of bins and the number of alphabet partitions at starting positions. Instances of FSIndex having many partitions at first few positions perform well (\texttt{SPEQ06}, \texttt{SPNA09}), those that have few partitions with many letters per partition, less so. 

Clearly, if a bin has a single letter partition at its first position, the distance at that position need be only retrieved once, at the start of the scan, independently of the number of fragments the bin contains. The effects for the second and subsequent positions are less prominent, if only for the reason that using many partitions would result in many bins being empty. The actual composition of the dataset is also important, as Figure \ref{fig:FSscale} (e) attests: although same partitions are used and \texttt{nr0288K} is almost twice as large, \texttt{SPEQ09} scans fewer characters. The possible reason lies in the nature of SwissProt, which, as a human curated database, is biased towards the well-researched sequences which are more related among themselves while not necessarily being representative of the set of all known proteins. On the other hand, \texttt{nr0288K} is a random sample from the \texttt{nr} database which is exactly the non-redundant set of all known proteins. 

In our evaluations, the average proportion of residues scanned per bin varies from 30\% (\texttt{SPEQ06}, length 6) to over 85\% (\texttt{nr018K}, length 9). The percentage of characters scanned grows slowly with increase of the number of neighbours retrieved, most probably because the number of bins accessed also grows, requiring that at least one full sequence is scanned.

To summarise, subindexing of bins does produce some savings, the exact amount depending on the dataset and alphabet partitioning. However, and this is further attested by poor performance of the suffix array compared to FSIndex (Table \ref{tbl:FScomp}), the good performance of FSIndex is mostly due to alphabet partitioning.

\subsection{Effect of similarity measures}

Table \ref{tbl:FSmatdata} indicates very little difference in performance of the same instance of FSIndex with respect to different similarity measures. This should not be a surprise because the BLOSUM matrices are indeed very similar, modeling the same phenomenon in slightly different ways but generally retaining the same groupings of amino acids. The PSSM-based searches also performed well, mainly because the PSSMs are usually constructed out of sets of sequences that are strongly conserved at least in one or two positions, and hence, in those positions, the `distances' to all other clusters are so large that many branches of the implicit search tree can be pruned. 

\subsection{Scalability}

Unlike both M-tree and mvp-tree, as well as most other known metric or spatial access methods \cite{CNBYM,HjSa03,SRF87}, FSIndex does not have a balanced tree. Therefore, the expected average and worst-case search time complexity is $O(n+K)$ -- the overhead is proportional to $K$, the number of inner nodes. Based on these considerations, it appears that FSIndex is not scalable for queries of a fixed radius. However, the performance can be to a large extent controlled by the choice of alphabet partitions and hence some scalability can be achieved by using more partitions for larger datasets in order to reduce the scanning time while incurring some additional overhead. 

Figure \ref{fig:FSscale} (b) indicates that FSIndex is scalable with respect to the number of nearest neighbours retrieved -- the number of residues needed to be scanned grows sublinearly with dataset size (in fact, the exponent is 0.25 to 0.3). The exponent for the growth of the number of scanned points (graphs not shown here) is about 0.4, indicating that using PATRICIA-like structure improves scalability. The most likely reason for sublinear growth of the number of items needed to be scanned is that the search radius needed to retrieve a fixed number of nearest neighbours decreases with dataset size, requiring fewer bin accesses.

\subsection{Comparison with other indexing schemes}

Results of Subsection \ref{subsec:FScomp} indicate that FSIndex decisively outperforms all other indexing schemes considered. M-tree performed the worst, needing to scan 1.3 million fragments of length 10 in order to retrieve the nearest neighbour. The performance of mvp-tree is not much better, taking into account the dimensionality: it requires scanning about 1 million fragments of length 9 to retrieve the nearest neighbour. Suffix array was generally performing better than mvp-tree, except for retrieving the nearest neighbour of length 6.

In the case of suffix arrays, it is clear that large alphabet and relatively small dataset are responsible for relatively poor performance. Also note that suffix trees (and hence suffix arrays) generally are not good approximations of the geometry with respect to $\ell_1$-type distances -- two fragments lacking a common prefix may have a small distance. It should be noted that performance of suffix array based scheme appears to improve with fragment length compared to FSIndex.

The poor performance of M-tree and mvp-tree is somewhat surprising because Mao, Xu, Singh and Miranker \cite{MXSM03} have recently proposed using exactly M-tree for fragment similarity searches. However, on closer inspection, several differences appear. First, Mao, Xu, Singh and Miranker use a different metric. More importantly, they use a significantly improved M-tree creation algorithms. Finally, if their results are compared with those from Figure \ref{fig:FMTNNperf} (this can be done at least approximately because the same fragment length was used and the size of the yeast proteome dataset used in \cite{MXSM03} was very close to the size of SwissProt sample used in our experiment), it appears that there is no more than 10-fold improvement. While this is quite significant, the total performance appears still worse than that of FSIndex. For more detailed comparisons it would be necessary to obtain the code of the improved M-tree from \cite{MXSM03} and run a full suite of comparison experiments.

An interesting possibility would be to try and combine our approach with the Multi Resolution String (MRS) index structure for string datasets proposed by Kahveci and Singh \cite{KahSingh:2001} in a general context and further adapted for protein database searches, cf. e.g. \cite{BCKSW04}. The MRS scheme is based on mapping a string dataset to a lower-dimensional space of sequences of integers by means of the first few Haar wavelet transform coefficients. In the context of protein sequence databases, the first wavelet coefficient corresponds to the well-known concept of the aminoacid content of a fragment, while the second coefficient is the difference between the aminoacid content vectors of the first and the second halves of a given fragment. The MRS scheme applies to string datasets in any possible alphabet, and does not seem to capture the intrinsic geometry of the alphabet itself. For this reason, it appears to us that this approach and ours are ``complimentary'' to each other, and combining them in the context of datasets of protein fragments could result in a further performance gain. 

\subsection{Range vs $k$NN queries}

Performance of the branch-and-bound algorithm depends on the order of nodes visited -- it is to a great advantage if the nodes containing data points closest to the query are visited first so that the bounding radius becomes small early on. A frequently used solution \cite{CPZ97,HjSa03} is to traverse the tree breadth-first, keeping the nodes to be visited in a second priority queue, where the priority of a node is given by the upper bound of the distance of its covering set from the query.

The second priority queue is not used for the FSIndex based $k$NN search. Since the implicit tree is heavily unbalanced, the branches with smallest depth are visited first with a similar effect without the overhead of the second priority queue. The visiting order of nodes is ensured in the outer loop of the \textsc{CheckNode} function where the index $j$ starts at $m-1$, decreasing to $i$ (Algorithm \ref{alg:FSchkbin}). Since the order does not affect the range search performance, the same code can be used for range search.

While queries based on more than one similarity measure can be used on a single FSIndex, it is to be expected that similarity measures different from the one originally used to determine the partitions would have worse performance. 

\subsection{Applications to general protein sequence search}

General protein sequence search involves retrieval of sequences based on gapped similarity measure. Retrieval based on sequential scan with respect to the Smith-Waterman local similarity score \cite{SW81}, the most commonly used similarity measure, is infeasible for large datasets and a number of heuristic search tools were proposed, the most famous being the NCBI BLAST \cite{altschul97gapped}.

To perform a database search, BLAST takes the query sequence and divides it into very short fragments (length 3 is common for proteins). It then searches for close matches of these fragments to the fragments of the database sequences. When a match is found, it attempts to extend this `seed' in order to produce a longer, gapped alignment.

Recently, there have been a number of attempts to use indexing schemes on DNA or protein fragments so as to increase the length of seeds and hence accelerate the search \cite{Buhler01,GiWaWaVo00,MXSM03,TCOT03}. Efficient performance of FSIndex also recommends it for such purpose, by taking seeds of length 6-12. By taking longer seeds, the total number of seeds can be reduced, leading to less time being spent extending seeds that are subsequently rejected. We leave the follow-up on this idea for a subsequent investigation.

\section{Conclusion}

The efforts of the data engineering community in constructing indexing schemes for similarity-based information retrieval have been largely concentrated either on generic indexing schemes which would work for an arbitrary metric (or normed) space \cite{CPZ97,CNBYM,BoOzs97,SRF87}, or else on  q-grams or suffix tree-based schemes for string similarity \cite{BuKi01,NaBY00,NaBYSuTa01,HuAtIr01,Hu04}. Metric space indexing schemes are often subject to the curse of dimensionality (as witnessed in a part of our present investigation), while the suffix tree-based schemes are very efficient but may suffer from large use of memory, although we are seeing good attempts towards removing this limitation \cite{HuAtIr01,Hu04}

The FSindex scheme proposed in our paper seems to explore the geometry of datasets of short protein fragments in a very fortunate way. We believe that the efficiency of the scheme can be further improved, but already in its present form it can be put to a number of uses. We intend to use FSindex in the first stage of searching the complete protein fragment datasets for functional motifs. We believe that alphabet reduction can be used in other indexing methods, in particular in large suffix trees \cite{Hu04} if some space reduction is desired at a very little loss in performance. Another promising application is to use FSindex for seeds in local similarity search tools. 

The proposed indexing scheme has a theoretical significance. In \cite{PeSt06}
we have developed a general approach to building new indexing schemes from
old by applying a number of fundamental operations, and explained how a
a prototype version of the FSIndex scheme can
be obtained in this way from the so-called projective reduction to a suffix tree. It is conceivable that very efficient indexing schemes can be engineered in the future in the same manner by properly combining a number of very simple tools. 

We believe that FSIndex is a successful realization of this idea. In particular, we have made public a new software tool for assigning biological function to protein fragments, called PFMFind, based on FSIndex as a low-level core component. The source code is available at \url{http://www.vuw.ac.nz/biodiscovery/Publications/PFMFind/index.aspx}.


\section*{Acknowledgements}
The authors are grateful to Bill Jordan for stimulating discussions. The investigation was supported by the University of Ottawa research funds. We especially thank The High Performance Computing Virtual Laboratory (HPCVL) which provided us with computational resources required to complete this investigation. The first named author (A.S.) was also supported by a Bright Future PhD scholarship awarded by the NZ Foundation for Research, Science and Technology jointly with the Fonterra Research Centre and by Victoria University of Wellington research funds. We also thank Marco Patella for providing us with the code for mvp-tree. Comments of two anonymous referees which have led to a large number of improvements are much appreciated. 

\def\cdprime{$''$}


\end{document}